\documentclass[prd,twocolumn,amsmath,amssymb,floatfix,superscriptaddress,nofootinbib]{revtex4-1}

\usepackage{graphicx}
\usepackage{amssymb}
\usepackage{amsmath}
\usepackage{bm}
\usepackage{color}

\DeclareMathAlphabet\mathbfcal{OMS}{cmsy}{b}{n}
\definecolor{darkgreen}{RGB}{50,150,0}
\definecolor{purple}{cmyk}{0.5,1.0,0,0}

\DeclareMathOperator{\Tr}{Tr}

\def\edth{\;\raise1.0pt\hbox{$'$}\hskip-6pt\partial}
\def\baredth{\;\overline{\raise1.0pt\hbox{$'$}\hskip-6pt
\partial}}

\newcommand{\vh}[1]{\ensuremath{\hat{\bm{#1}}}}
\newcommand{\vp}{\bar{\vc{m}}}
\newcommand{\vm}{\vc{m}}
\newcommand{\vc}[1]{\ensuremath{{#1}}}

\newcommand{\f}[2]{\frac{#1}{#2}}  
\newcommand{\mk}[1]{\left( #1 \right)}  
\newcommand{\kk}[1]{\left[ #1 \right]}

\newcommand{\stucky}{{St\"{u}ckelberg} }

\newcommand{\tr}[1]{[#1]}
\newcommand{\dd}{d}  
\newcommand{\ul}[3]{#1^{#2}_{ \hphantom{#2}\! #3}}
\newcommand{\lu}[3]{#1_{#2}^{ \hphantom{#2}\! #3}}
\newcommand{\bgamma}{\boldsymbol{\gamma}}
\newcommand{\boldeta}{\boldsymbol{\eta}}
\newcommand{\bOurG}{{\chi}}
\newcommand{\OurG}{\bar \chi}
\DeclareMathOperator{\diag}{diag}
\def\be{\begin{equation}}
\def\ee{\end{equation}}
\def\ben{\begin{equation} \nonumber}
\def\een{\end{equation}}
\def\ban{\begin{eqnarray*}}
\def\ean{\end{eqnarray*}}
\def\ba{\begin{eqnarray}}
\def\ea{\end{eqnarray}}
\def\({\left(}
\def\){\right)}

\newcommand{\Mpl}{M_{\rm Pl}}
\newcommand{\Amat}{\mathbb{A}}
\newcommand{\Bmat}{\mathbb{B}}
\newcommand{\Cmat}{\mathbb{C}}
\newcommand{\Pmat}{\mathbb{P}}
\newcommand{\Qmat}{\mathbb{Q}}
\newcommand{\Lmat}{\mathbb{L}}
\newcommand{\Imat}{\mathbb{I}}
\newcommand{\Jmat}{\mathbb{J}}
\newcommand{\Nmat}{\mathbb{N}}
\newcommand{\Rmat}{\mathbb{R}}
\newcommand{\Smat}{\mathbb{S}}
\newcommand{\qa}{q} 
\newcommand{\Hv}{q}
\newcommand{\Leff}{\Lambda}
\newcommand{\Lnorm}{A}
\newcommand{\uvec}{{\bf u}}
\newcommand{\vvec}{{\bf v}}
\newcommand{\smetric}{\sigma}

\newcommand{\Comment}[1]{{}}

\definecolor{ultramarine}{rgb}{0.07, 0.04, 0.56}
\definecolor{cadmiumgreen}{rgb}{0.0, 0.42, 0.24}
\definecolor{indigo(dye)}{rgb}{0.0, 0.25, 0.42}
\usepackage[linktocpage=true]{hyperref}
\hypersetup{
colorlinks=true,
citecolor=ultramarine,
linkcolor=cadmiumgreen,
urlcolor=indigo(dye),
pdfauthor={},
pdftitle={},
pdfsubject={}
}

\begin{document}

\title{Self-accelerating Massive Gravity: \\ Hidden Constraints and  Characteristics}
\author{Pavel Motloch}
\affiliation{Kavli Institute for Cosmological Physics, Department of Physics, University of Chicago, Chicago, Illinois 60637, U.S.A}
\author{Wayne Hu}
\affiliation{Kavli Institute for Cosmological Physics, Department of Astronomy \& Astrophysics,  Enrico Fermi Institute, University of Chicago, Chicago, Illinois 60637, U.S.A}
\author{Hayato Motohashi}
\affiliation{Kavli Institute for Cosmological Physics, Department of Astronomy \& Astrophysics,  Enrico Fermi Institute, University of Chicago, Chicago, Illinois 60637, U.S.A}

\begin{abstract}
\noindent
Self-accelerating backgrounds in massive gravity provide an arena to explore the Cauchy problem for derivatively coupled fields that obey complex constraints which reduce the phase
space degrees of freedom.  
We present here an algorithm based on the Kronecker form of a matrix pencil that finds all hidden constraints, for example those
associated with derivatives of the equations of motion, and  characteristic curves for
any 1+1
dimensional system of linear partial differential equations.   With the Regge-Wheeler-Zerilli decomposition of metric perturbations  into
angular momentum and parity states, this technique applies to  fully 3+1 dimensional
perturbations of massive gravity around any isotropic self-accelerating background. 
Five spin modes of the massive graviton propagate once the constraints are imposed:
two spin-2 modes  with luminal characteristics present in the massless theory as well as
two spin-1 modes and one spin-0 mode.   Although the new modes all possess the same ---
typically spacelike ---
characteristic curves, the spin-1 modes are parabolic while the spin-0 modes are hyperbolic. 
The joint system, which remains coupled by non-derivative terms, cannot be solved as a simple Cauchy problem from a single non-characteristic surface.    We also illustrate the generality of the algorithm with other cases where derivative constraints reduce the number of propagating degrees of freedom or order of the equations.
\end{abstract}

\maketitle

\section{Introduction}

Using an auxiliary flat fiducial metric, de~Rham, Gabadadze and Tolley (dRGT) first constructed a consistent interacting theory of a massive spin-2 graviton~\cite{deRham:2010kj}.   This theory possesses a class of self-accelerating cosmological solutions where the massive graviton
potential plays the role of a cosmological constant
\cite{deRham:2010tw,Koyama:2011xz,Gumrukcuoglu:2011ew,Koyama:2011yg,Nieuwenhuizen:2011sq,
Berezhiani:2011mt,D'Amico:2011jj,Gratia:2012wt,Kobayashi:2012fz,
Volkov:2012cf,Volkov:2012zb,Motohashi:2012jd,Gratia:2013uza}.

The behavior of perturbations around these cosmological solutions is not {fully} understood. Initially,
there appeared to be several inconsistencies like coordinate dependence of the number of
propagating degrees of freedom \cite{Khosravi:2013axa} and related claims of existence of
strong coupling around particular solutions \cite{Gumrukcuoglu:2011zh}. These were
shown to be related to the existence of superluminally
propagating modes \cite{Motloch:2015gta}, which are a typical feature of isotropic perturbations around these
cosmological solutions. For a particular solution where
both the spacetime and fiducial metric are manifestly 
homogeneous \cite{Gumrukcuoglu:2011zh}, anisotropic modes are superluminal as well.    The Hamiltonian for the
isotropic modes around any self-accelerating cosmological solution is also unbounded from below
\cite{Khosravi:2013axa}.   Finally, on specifically constructed alternate backgrounds,
perturbation characteristics have also been shown to be superluminal \cite{Deser:2012qx,Deser:2013eua,Deser:2014hga,Deser:2014fta,Deser:2015wta}.

In this paper we investigate the behavior of all linear metric perturbations around a general
self-accelerating vacuum dRGT solution completing the analysis of Ref.~\cite{Motloch:2015gta}. As a typical solution lacks translation invariance,
it is not possible to employ the standard scalar-vector-tensor decomposition. However, thanks to the rotational invariance of the background and parity invariance of the theory, it is possible to decouple 
the system into angular momentum and parity states using the Regge,
Wheeler~\cite{Regge:1957td}, and Zerilli~\cite{Zerilli:1970se} formalism.   This formalism
was originally developed to study  perturbations around the Schwarzchild metric in 
general relativity (see  \cite{DeFelice:2011ka, Motohashi:2011pw, Motohashi:2011ds, Kobayashi:2012kh,
Kobayashi:2014wsa, Ogawa:2015pea} for extensions in modified gravity theories).

For a given angular momentum and parity, the various components of the metric perturbations are 
derivatively and non-derivatively coupled in a complicated constrained structure that reflects
the fact that only 5 spin modes of the massive graviton propagate.  We present here an algorithm capable of 
finding hidden constraints and characteristic curves for any set of linear partial differential equations
in 1+1 dimensions, which has
application beyond dRGT. Using this algorithm, we determine the characteristic
curves for all dRGT modes and identify their hyperbolic, parabolic or elliptic nature  
for their potential joint solution from initial data. 

The paper is organized as follows. In \S\ref{sec:dRGTIntro} we review the construction of
self-accelerated background solutions~\cite{Gratia:2012wt}, perturbation Lagrangian
around them~\cite{Motloch:2014nwa} and example vacuum solutions. In \S\ref{sec:methodology} we review the
Regge-Wheeler-Zerilli analysis, and provide an  summary of the algorithm we use
for finding the characteristics.   The Appendices contain a full explanation of the
algorithm (\S\ref{sec:AlgorithmAppendix}), decomposition techniques (\S\ref{sec:decomposition}) 
and crosscheck using an alternate method of auxiliary variables (\S\ref{ssec:oddal}).
In \S\ref{sec:Odd} and \S\ref{sec:Even} we then 
investigate odd and even parity perturbations around dRGT cosmological solutions.   
We discuss these results in \S\ref{sec:discuss}.

\vfill

\section{Self-accelerating solutions in massive gravity}
\label{sec:dRGTIntro}

In this section we provide a concise review of  the dRGT theory (\S\ref{sec:dRGT}), 
its
self-accelerating isotopic background solutions (\S\ref{sec:isotropic}), perturbations
in unitary gauge (\S\ref{sec:perts}), and specific vacuum background solutions (\S\ref{sec:vacuum}).

\subsection{dRGT theory}
\label{sec:dRGT}

The Lagrangian density for the dRGT \cite{deRham:2010kj} nonlinear theory of a massive spin-2 graviton
is given by: 
\be
\label{drgt}
{\cal L} = \sqrt{-g}\frac{M_{\rm Pl}^2}{2}\left[ R-m^2\sum_{k=0}^4 \frac{\beta_k}{k!} F_k\left(\bgamma\right)
\right],
\ee
where $M_{\rm Pl}^2 = (8\pi G)^{-1}$ is the reduced Planck mass,
\begin{align}
F_0(\bgamma) & = 1, \nonumber\\
F_1(\bgamma) & = \tr{\bgamma}, \nonumber\\
F_2(\bgamma) & =  \tr{\bgamma}^2 - \tr{\bgamma^2} , \\
F_3(\bgamma) & =\tr{\bgamma}^3 - 3 \tr{\bgamma} \tr{\bgamma^2} + 2 \tr{\bgamma^3} , \nonumber\\
F_4(\bgamma) &= \tr{\bgamma}^4 - 6 \tr{\bgamma}^2 \tr{\bgamma^2} + 3 \tr{\bgamma^2}^2 + 8 \tr{\bgamma} \tr{\bgamma^3}
- 6 \tr{\bgamma^4} ,
\nonumber
\end{align}
and $[\,]$ denotes the trace of the enclosed matrix. The matrix $\bgamma$ is the  square root of the  product of the inverse spacetime metric ${\bf g}^{-1}$ and a flat fiducial metric $\boldsymbol{\Sigma}$
\be
\ul{\gamma}{\mu}{\alpha} \ul{\gamma}{\alpha}{\nu} = g^{\mu\alpha}\Sigma_{\alpha\nu} .
\label{eqn:gamma}
\ee
$\boldsymbol{\Sigma}$
is itself related to the standard Minkowski metric $\boldeta$ via a coordinate transformation
using \stucky scalars $\phi^A$ 
\be
 \Sigma_{\mu\nu} = \eta_{A B} \partial_\mu\phi^A\partial_\nu\phi^B ,
\ee
which restores diffeomorphism invariance to the theory.  Where this transformation is not invertible 
the dRGT degrees of freedom encounter a determinant singularity \cite{Gratia:2013gka}.  
Smooth continuation of
solutions on the other side of a determinant singularity is sometimes but not always possible 
\cite{Motloch:2015gta}. 

The parameters of the dRGT theory  are  $\{\alpha_3,\alpha_4\}$, which
control the $\beta_k$ through
\begin{align}
\beta_0 &= -12 (1+ 2\alpha_3+2\alpha_4), \nonumber\\
\beta_1 &= 6(1 + 3 \alpha_3 + 4\alpha_4),\nonumber\\
\beta_2 &= -2(1+ 6 \alpha_3+12\alpha_4 ), \\
\beta_3 &= 6(\alpha_3+ 4\alpha_4), \nonumber\\
\beta_4 &= -24 \alpha_4,\nonumber
\end{align}
and the graviton mass $m$.

\subsection{Isotropic background solutions}
\label{sec:isotropic}

The dRGT theory possesses 
solutions for
any  isotropic spacetime metric \cite{Gratia:2012wt}
where the stress-energy associated with the graviton potential in Eq.~\eqref{drgt} behaves as a cosmological constant.  

Given an isotropic line element,
\be
	{\dd}s^2 = -b^2(t,r) \dd t^2 + a^2(t,r) \big(\dd r^2 + r^2  \dd \Omega_2^2 	\big),
	\label{isotropicmetric}
\ee
where $\dd \Omega_2^2$ is the line element on a 2-sphere, and isotropic 
\stucky fields,
\ba
	 \phi^0 &=& f(t,r),\nonumber\\
	 \phi^i &=& g(t,r) \frac{x^i}{r} ,
	\label{stuckyback}
\ea
this class of self-accelerating solutions requires
\ba
	g(t,r) =x_0  a(t,r)r.
	\label{eqn:gsoln}
\ea
The constant $x_0$ solves the polynomial equation $P_1(x_0)=0$ with
\be
P_1(x) \equiv 2(3-2x)+6(x-1)(x-3)\alpha_3+24(x-1)^2\alpha_4.
\ee
Distinct self-accelerating \stucky backgrounds represent different solutions of 
\begin{equation}
\sqrt{X} = \frac{W}{x_0}+x_0,
\label{eqn:feqnsoln}
\end{equation}
where
\begin{align}
X & \equiv\Bigl(\frac{\dot{f}}{b}+\mu\frac{g'}{a}\Bigr)^2-\Bigl(\frac{\dot{g}}{b}+\mu\frac{f'}{a}\Bigr)^2, \nonumber\\
W & \equiv \frac{\mu}{ab} \( \dot f g' - \dot g f' \),
\label{eqn:XW}
\end{align}
with branches due to the matrix square root $\bgamma$ defined in Eq.~\eqref{eqn:gamma}
allowing $\mu\equiv \pm 1$.   Here and throughout, we choose $\mu=1$ and overdots denote
derivatives with respect to $t$ whereas primes denote derivatives with respect to $r$.
Where $W=\pm\infty, 0$ or is undefined because either $f$ or $g$ are not
continuously differentiable, there exists a determinant singularity \cite{Gratia:2013gka}.

For any such solution the effective stress tensor due to the presence of the non-derivative graviton interactions takes the form of an effective cosmological constant
\be
\label{effTmunu}
T_{\mu\nu} = - \Leff M_{\rm Pl}^2 g_{\mu\nu},
\ee 
where
\be
\Leff =  \frac{1}{2} m^2 P_0(x_0),
\ee
with 
\begin{align}
P_0(x) &= - 12 - 2 x(x-6) - 12(x-1)(x-2)\alpha_3 
\nonumber\\&\qquad -24(x-1)^2\alpha_4 ,
\end{align}
defining its dependence on dRGT parameters.

\subsection{Perturbation Lagrangian}
\label{sec:perts}

Ref.~\cite{Motloch:2014nwa} derived the covariant form for the  quadratic Lagrangian for perturbations 
around the isotropic self-accelerating solutions of the previous section.  Here we
consider a specific gauge for the perturbations, called unitary gauge,  in which the
\stucky perturbations vanish
\be
	\phi^{A} = \bar \phi^{A} ,
\ee
where bar denotes the background quantity.  This can always be accomplished by an infinitesimal
 transformation $x^\mu \rightarrow x^\mu + \xi^\mu$ which changes these scalar fields by
\begin{equation}
\delta \phi^A = \frac{\partial \bar\phi^A}{\partial x^\mu} \xi^\mu.
\label{eqn:unitary}
\end{equation}
Inverting this relation, a \stucky fluctuation $\delta \phi^A$
can always be gauged away fixing $\xi^\mu$ entirely
as long as $\partial \bar\phi^A/\partial x^\mu$ is not
singular, i.e.\ away from a determinant singularity.  If the background solution can be continued on the other side
of a determinant singularity, a new unitary gauge can be established there as well.

Notice the background \stucky fields are in general nonzero and so this unitary condition refers  to  the fact that the perturbed degrees of freedom propagating on the background
come only from the metric 
\be
	g_{\mu\nu} = \bar g_{\mu\nu}+
	h_{\mu\nu} .
\ee
The quadratic Lagrangian for the metric fluctuations $h_{\mu\nu}$ is then
\cite{Motloch:2014nwa}
\ba
	\mathcal{L}_2  &=& \mathcal{L}_{hh}^{\rm (EH)}+\mathcal{L}_{hh}^{(\Lambda)} +\Lnorm \sqrt{-\bar g} \Mpl^2  B^{\mu\nu\alpha\beta} h_{\mu\nu} h_{\alpha\beta},
\label{SimplifiedLagr}
\ea
where the Einstein-Hilbert piece 
\ba
\label{EinsteinHilbert}
\frac{ \mathcal{L}_{hh}^{\rm  (EH)} }{\sqrt{-\bar g} \Mpl^2 }&=&\left( \frac{1}{2}
h^{\mu\alpha}\lu{h}{\alpha}{\nu} - \frac{1}{4}  h h^{\mu\nu} \right) \bar R_{\mu\nu}
\\
&& +
 \left(\frac{1}{16}  h^2 - \frac{1}{8} h_{\mu\nu} h^{\mu\nu} \right) \bar R -\frac{1}{8} h^{\mu\nu;\alpha} h_{\mu\nu;\alpha} \nonumber\\
&&
 + \frac{1}{4} h^{\mu\nu;\alpha} h_{\nu\alpha;\mu}
+\frac{1}{8} h_{;\alpha} h^{;\alpha} 
-\frac{1}{4} \ul{h}{\mu\nu}{;\nu} h_{;\mu} \nonumber,
\ea
the effective background cosmological constant piece
\ba
\frac{\mathcal{L}^{(\Lambda)}_{hh}}{\sqrt{-\bar g}\Mpl^2} &=&  \left(\frac{1}{4} h_{\mu\nu}
h^{\mu\nu}  -\frac{1}{8}  h^2  \right)\Lambda ,
\ea
and the dRGT potential piece
\ba
	B^{\mu\nu\alpha\beta}&=&
		\frac{ [\bar\bOurG]}{8}\( \bar g^{\mu\nu} \bar g^{\alpha \beta}-\frac{1}{2} \bar g^{\mu\beta}\bar g^{\nu\alpha} - \frac{1}{2} \bar g^{\mu \alpha} \bar g^{\nu \beta} \)
	\nonumber\\
	&&+ \frac{1}{16} \(\bar g^{\mu\alpha} \OurG^{\nu \beta} + \bar g^{\nu\beta} \OurG^{\mu\alpha} + \bar g^{\mu \beta} \OurG^{\nu\alpha} + \bar g^{\nu \alpha} \OurG^{\mu\beta}\)
	\nonumber\\
	&&- \frac{1}{8}\(\bar g^{\mu\nu}\OurG^{\alpha\beta} + \bar g^{\alpha\beta}
	\OurG^{\mu\nu}\) ,
\ea
with the normalization
\begin{equation}
\Lnorm= \frac{x_0^2 P_1'(x_0)}{4}m^2 .
\end{equation}
Here $\bar R_{\mu\nu}$ is the usual Ricci tensor built out of the background metric,
$\bar R$ is its trace, and
\be
\label{chi}
	\OurG_{\mu\nu} = \frac{1}{x_0} \bar \gamma_{\mu\nu} - \bar g_{\mu\nu} ,
\ee
whose only nonzero components are 
\ba
	\OurG_{11} &=& \frac{ x_0^2  b^2 -  \dot{  f}^2 +  \dot{  g}^2}{x_0^2  +
	 W} , \nonumber\\
	\OurG_{12} &=& \frac{ (\dot { g}  g' - \dot { f}  f')}{x_0^2 +  W} , \nonumber\\
	\OurG_{22} &=& - \frac{ x_0^2  a^2 +   f'^2 -   g'^2}{x_0^2 +  W} .
\ea
From the background equations of motion (EOMs) it can be shown that they satisfy
\be
\label{BackgroundEOM}
	\OurG_{11} \OurG_{22} = \OurG_{12}^2 .
\ee
To quadratic order, all covariant derivatives can be taken with respect to 
the background metric.

\subsection{Vacuum solutions}
\label{sec:vacuum}

In the absence of matter, the effective cosmological constant for the background solution leads
a de Sitter spacetime  with an expansion rate $H=\sqrt{\Leff/3}$.
Closed isotropic coordinates 
\begin{equation}
\label{eqn:closeddS}
\dd s^2 = -\dd t^2  +\left[ \frac{ \cosh{(H t)} }{1 +(H r)^2/4} \right]^2 \left(\dd r^2 +
r^2 \dd\Omega_2^2\right) ,
\end{equation}
chart the entire spacetime, where  $t \in (-\infty, \infty)$ and $r\in [0, \infty)$. 
For the purposes of illuminating the causal structure of solutions using conformal diagrams,
it is also useful to introduce  conformal coordinates
\begin{equation}
\label{conformal_metric}
\dd s^2 = \left( \frac{1}{H \sin \eta}\right)^2 \left(-\dd\eta^2 + \dd\chi^2 + \sin^2\chi \dd\Omega_2^2\right),
\end{equation}
where 
\begin{eqnarray}
 \sinh (H t) &=& - \cot \eta  ,\nonumber\\
 H r &=& 2 \tan(\chi/2) ,
 \label{eqn:closedtr}
\end{eqnarray}
restricts $\eta \in (0, \pi)$ and $\chi \in [0, \pi]$.

Specific solutions are defined by the background temporal \stucky field $f$.  
Although our treatment is fully general, we will
illustrate our results using two classes of solutions, the so-called 
``open solution'' of \cite{Gumrukcuoglu:2011ew}
\be
	\label{solMukohyama}
	f= f_o(\eta,\chi) = \frac{x_0}{H} \cot\eta,
\ee
and the family of solutions from Ref.~\cite{Koyama:2011yg} 
\ba
	\label{solKoyama}
	f=f_C(\eta,\chi) &=& \frac{x_0}{C H} \left(    \ln \left\lvert \frac{C^2 (\cos \chi + \cos
	\eta)}{\sin\eta(1-y)}\right\rvert - y \right),\nonumber\\
	y &=& \sqrt{1 + C^2( \sin^2\chi/\sin^2\eta -1)},
\ea
where $C \in (0, 1]$ is a free parameter and $y \in [0,\infty)$.  Properties of these background solutions
including their determinant singularities were extensively discussed in Ref.~\cite{Motloch:2015gta}.

\section{Methodology}
\label{sec:methodology}

Here we present our main analysis techniques.  In \S\ref{sec:RW}, we decompose metric fluctuations into parity, angular momentum and spin components using the harmonic functions for tensors on the 2-sphere reviewed in \S\ref{sec:harmonics}.   Parity and 
angular momentum states obey decoupled EOMs as discussed in \S\ref{sec:eomchi}.
In \S\ref{sec:charsummary}, we show how to resolve hidden 
constraints and determine the characteristics and appropriate boundary conditions for
derivatively coupled systems like the metric modes of dRGT.   Details of this general
algorithm,  which uses the Kronecker decomposition of a matrix pencil reviewed in
\S\ref{sec:KroneckerAppendix}, are given with pedagogical examples in Appendix \ref{sec:AlgorithmAppendix}.

\subsection{Regge-Wheeler-Zerilli  decomposition}
\label{sec:RW}

The analysis of metric perturbations around isotropic dRGT vacuum solutions  is more complicated than standard cosmological perturbation theory due to the background
\stucky fields or more specifically the presence of a new background tensor $\OurG_{\mu\nu}$ in addition to the homogeneous and isotropic
metric $\bar g_{\mu\nu}$.  In this case the normal modes of fluctuations are no longer eigenfunctions of
three dimensional Laplacian and the usual decoupling of scalar, vector and tensor fluctuations does not apply.

While the dRGT background is generally no longer translationally invariant, it remains rotationally invariant and so the normal modes
are characterized by their angular momentum.   In addition, the quadratic Lagrangian is parity invariant and so
the even and odd parity modes are decoupled.   The analysis of metric perturbations under these conditions 
follows  the Regge-Wheeler-Zerilli (RWZ) analysis \cite{Regge:1957td,Zerilli:1970se} originally introduced 
for the similarly inhomogeneous Schwarzschild metric.

Here the 10 metric fluctuations of the symmetric $h_{\mu\nu}$ are decomposed in spherical coordinates
$\{t,r,\theta,\phi \}$ and classified according to their transformation properties under rotation and parity.  
This classification is reviewed in \S\ref{sec:harmonics} and implies the presence of conserved
quantum numbers for the total angular momentum $\ell$, azimuthal angular momentum $m$, and parity $E$ for even
and $B$ for odd, 
as well as the spin $s$ of the various components:
\ba
\label{RWExpansion}
h_{tt}&=& H_{0}^{\ell m}\, Y_{\ell m},\quad   
h_{tr}=H_{1}^{\ell m}Y_{\ell m}, \quad  
h_{rr}=H_{2}^{\ell m} Y_{\ell m},\nonumber\\
h_{ta}&=& h_{0}^{\ell m} Y^B_{\ell m,a}
 + \beta^{\ell m}Y^E_{\ell m,a},
\nonumber\\
h_{ra}&=& h_{1}^{\ell m}Y^B_{\ell m,a}
 + \alpha^{\ell m}Y^E_{\ell m,a},\nonumber\\ 
h_{ab}&=& h_{2}^{\ell m}Y^B_{\ell m,ab} + 
 G^{\ell m} Y^E_{\ell m,ab} + K^{\ell m}  Y_{\ell m}\smetric_{ab} , \nonumber
\ea 
where $a,b \in \{ \theta,\phi \}$, $\smetric_{ab}$ is the metric of the 2 sphere and the summation over $\ell,m$ is implicit.
Here $\mathbb{Y}_{\ell m}^{X}$ with $X \in \{ E,B\}$ are the tensor spherical harmonics and depend on $\{ \theta,\phi \}$
whereas the fields or coefficients whose spin and parity are summarized in Table~\ref{tab:Fields} are functions of 
$\{ t,r \}$.  Note that the angular momentum states for a given spin $s$ are restricted to $\ell \ge s$.
We differ slightly from the original RWZ analysis by  removing the spin-0 or trace piece of the rank-2 angular tensors
which better isolates their rotational properties.

\begin{center}
\begin{table}
\caption{Metric Modes.  Here $a,b \in \{ \theta,\phi\}$.}
\begin{tabular}{cccccccc}
\hline\hline
``$E$" even \quad &$H_0$ & $H_1$ & $H_2$ & $K$  & $\beta$ & $\alpha$    &$G$ \\\hline
spin & 0 & 0 & 0 & 0 & 1 & 1 & 2 \\
$\mu\nu$ & $tt$ & $tr$ & $rr$ & $ab$ & $ta$ & $ra$ & $ab$ \\
\hline 
\\
\hline\hline
``$B$" odd &--&--&--&--& $h_0$ & $h_1$ & $h_2$ \\\hline
spin &--&--&--&--& 1 & 1 & 2\\
$\mu\nu$ & --  & --  & --  & --  & $ta$ & $ra$ & $ab$\\
\hline\hline
\end{tabular}
\label{tab:Fields}
\end{table}
\end{center}

The symmetries of the background imply that groups of $(X,\ell, m)$ modes decouple and can be analyzed
independently.   More specifically given the quadratic Lagrangian density, we can immediately integrate over angles to decouple
the Lagrangian density in $\{ t, r\}$ into a sum over independent terms
\begin{eqnarray}
 \int d\theta d\phi {\cal L}_2 \equiv  \Mpl^2 \sum_{X \ell m } \mathcal{L}_{X,\ell m}
\end{eqnarray}
with the help of the orthogonality relations in Eqs.~(\ref{eqn:orthonormality}) and (\ref{eqn:angularidentities}).   Note that ${\cal L}_2$ involves covariant derivatives
on the 2 sphere and so different spin states of a given $(X,\ell, m)$ are coupled.

\subsection{Equations of motion and singular points}
\label{sec:eomchi}

From the quadratic Lagrangian for each set of $(X,\ell, m)$ modes, we can derive the coupled
EOMs as usual.  Isotropy of the background requires that the EOMs for all $m$
modes of a given $\ell$ and $X$ are the same.  
The fields for $m \ne 0$ are complex, but we
will use as the shorthand convention $h_1^2$ for $|h_1|^2$, etc and suppress  subscripts
$\ell$ and $m$ on the $E$ and $B$ Lagrangian densities and field variables from here on.

The spin components of a given angular momentum and parity obey 
a rather complicated set of coupled EOMs.
Although  kinetic terms for these modes come from the Einstein-Hilbert term, not the dRGT potential term,
the nature of the constraints differs crucially from general relativity.    The lapse and shift perturbations
$H_0, H_1, \beta, h_0$ are still non-dynamical but their elimination becomes more complicated.
Furthermore, there is no remaining gauge freedom in dRGT which in general relativity
eliminates 4 more variables.  
Naively, this would leave the dRGT modes with 6 remaining degrees of freedom rather than
the 2 of general relativity.   However the special Boulware-Deser ghost-free structure of
the dRGT Lagrangian eliminates the 6th mode leaving 5 remaining degrees of freedom 
to represent the  spin states of the massive graviton.   In the usual convention for
the polarization states these would be spin $2$:
even $G$ and odd $h_2$;  spin-1: even $\alpha$ and odd $h_1$; spin-$0$: even $K$ (with $H_2$ present but obeying a constraint).

We can make one further simplification to the EOMs by using 
the property of the background solution \eqref{BackgroundEOM} to eliminate 
\be 
	\OurG_{22} = \frac{\OurG_{12}^2}{\OurG_{11}} .
	\label{eqn:replacement}
\ee 
The disadvantage of this approach is that it cannot be applied at points
where $\OurG_{11} = 0$. However if we encounter such a point, it is generally possible
to switch the chart of the background to pass through it as  $\OurG_{11}$ is a component of a tensor not a
scalar.
Since our conclusions will be about coordinate invariant quantities such as
characteristic curves, they are then valid at all such spacetime points. It suffices to
show that we can find a chart where $\OurG_{12} \neq 0$ since this implies $\OurG_{11}
\neq 0$ through Eq.~(\ref{eqn:replacement}). We checked that it is not possible to have
$\OurG_{12} = 0$ in both closed isotropic slicing \eqref{eqn:closeddS} and flat isotropic
slicing anywhere besides $\eta = -\pi/2, \chi = \pi/2$. In closed slicing at this
point $\dot { g} = g' = 0$ and is thus a coordinate-invariant 
determinant singularity.  Here  the background
solution itself is undefined. At each point where the background
solution is defined, our analysis as presented thus works in at least one background coordinate frame.
At points where $\OurG_{11}=0$, it is actually still possible to implement our 
analysis directly, but the details would differ from those
presented below (see \S\ref{ssec:oddl2}).

\subsection{Constraints and characteristics}
\label{sec:charsummary}

As discussed in the previous section, the derivatively coupled EOMs for
the spin states of a given angular momentum and parity obey a complicated constraint
structure. These take the form of  differential
equations and  not algebraic relations with which the variables involved may be simply eliminated.   
In some cases it is still possible to employ techniques
involving auxiliary variables, but 
these must be introduced on a case by case basis and do not form a systematic means of proceeding.

For the purposes of counting degrees of freedom and 
investigating the characteristics along which field information propagates, we use a systematic 
method introduced in the Appendix~\ref{sec:AlgorithmAppendix} based on augmented first order EOMs.
A summary of the method is as follows:

\begin{enumerate}
\item  Reduce all EOMs to first order form by introducing auxiliary variables, e.g.~$u_t  = \partial u/\partial t$ along with these defining equations as additional EOMs.  Cast the
EOMs in a matrix form as 
\begin{equation}
\Amat \dot \uvec+ \Bmat \uvec ' + \Cmat \uvec = 0.
\label{eqn:vecform}
\end{equation}

\item Identify and complete the ``regular blocks" of these equations, which  evolve
$\uvec$ uniquely and consistently,  by incorporating hidden algebraic and derivative constraints.

\begin{enumerate}
\item If  $\Amat + \lambda \Bmat$ is invertible for some choice of $\lambda$ then Eq.~(\ref{eqn:vecform}) specifies the evolution of $\uvec$ in some suitable temporal coordinate.  $\Amat + \lambda \Bmat$ defines a regular pencil or block. Proceed to step 3. 

\item If $\Amat + \lambda \Bmat$ is singular, in addition to regular $\mu\times \mu$ blocks $\Rmat_\mu$,
it contains overdetermined $(\mu+1)\times \mu$ blocks $\Lmat^P_\mu$, 
underdetermined $\mu \times (\mu+1)$ blocks $\Lmat_\mu$  or both in its Kronecker decomposition (see \S\ref{sec:KroneckerAppendix}).  Eliminate redundancies and add all missing algebraic
and derivative constraints from the overdetermined blocks to the EOMs.  If constraints turn
all underdetermined blocks to regular blocks, proceed to step 3.
 
\item 
If underdetermined blocks remain then solutions are not unique, often due to gauge freedom which can be fixed by addition of
gauge constraints.  Add these as EOMs and repeat the previous step.

\end{enumerate}

\item Cast the regular blocks in Weierstrass  form $\Rmat_{\mu}(\Omega)$ and read off characteristics from their eigenvalues $\Omega$.   
Derivative blocks operate on some linear combination of original fields $\vvec = \Qmat^{-1} \uvec$.
If a characteristic is real and the degeneracy or dimension of a  block is 1, the block is hyperbolic;
if higher than  1, parabolic.  If a characteristic is complex, the block is elliptic.  If all blocks are hyperbolic, then the whole system is hyperbolic.

\end{enumerate}

The classification of the system as hyperbolic, parabolic or elliptic has direct implications for the type of initial or boundary data required.  In a hyperbolic system, fields are uniquely specified by the EOMs along 
characteristics, given  data on a surface that intersects the characteristics. If this surface is spacelike, then one
solves a Cauchy problem for the evolution of fields from initial conditions. For a coupled system,  a well-posed Cauchy problem requires a joint  surface that intersects all characteristics. The slope of the characteristic defines
the analog of lightcones, i.e.~the domain of dependence and influence.  Characteristics of a hyperbolic system
also define curves across which the EOMs do not specify the field evolution.
Hence field discontinuities can occur on characteristics, if they occur in the initial data, and their speed of propagation is
given by the slope.  We will call characteristics ``superluminal" whenever
they are spacelike.
Of course actual discontinuities would be beyond the regime of validity of dRGT as an effective theory.
For an alternative discussion
of related issues, see \cite{Deser:2014fta, Babichev:2016hys}.

In a parabolic system, the EOMs contain derivatives in the direction orthogonal to the
characteristics which carries information across them. The
prototypical example is the heat diffusion equation where the characteristics are constant
time surfaces. Since the domain of dependence then involves all of the characteristics
``upstream" from a given characteristic, usually one specifies consistent field data on a
given ``initial" characteristic and marches forward across the ``downstream"
characteristics or domain of influence. The domain of dependence also spans the extent of
the characteristic, which is typically spacelike, and so requires spatial boundary
conditions as well.

In an elliptic system, no real characteristics exist and so the domain
of dependence is the entire spacetime. Elliptic systems cannot be solved by marching
initial data forward with the EOMs.

The characteristic analysis is therefore a tool to study the nature of the boundary value problem  in the classical theory.
It is a precursor to solving for field configurations either analytically or numerically.
If and only if all subsystems are hyperbolic and share
a joint non-characteristic surface, can the Cauchy problem for the system as a whole
be solved from data on this surface.

\section{Odd ``B" modes}
\label{sec:Odd}

We begin the analysis of the propagating degrees of freedom, constraints and characteristics 
of metric fluctuations around vacuum self-accelerating dRGT solutions with the odd parity modes.
Odd parity modes are simpler due to the smaller number of degrees of freedom associated
with them.   
They also provide a useful cross check on our general EOM-based technique since 
there is an alternate approach of introducing auxiliary fields into the Lagrangian,
which we explain in Appendix \ref{ssec:oddal}.

We first present the quadratic Lagrangian in \S\ref{ssec:oddLag} and explain how the method works for
the special case of $\ell = 1$ where only a single spin 1 mode propagates in \S\ref{ssec:oddl1}.
We study the general case $\ell \geq 2$ where there is an additional spin 2 mode
in \S\ref{ssec:oddl2}.  We compare this analysis to the alternate approach of \S\ref{ssec:oddal} 
in \S\ref{ssec:oddlalt}.  A summary of the regular blocks and characteristics of both the
odd and even modes  is given in
Table~\ref{tab:Characteristics}.

\subsection{Lagrangian}
\label{ssec:oddLag}

As discussed in \S\ref{sec:RW}, the normal mode decomposition of metric fluctuations decouples
the Lagrangian density into independent pieces for a given angular momentum $\{ \ell, m \}$ and parity state.
For each odd or $B$ set, the Lagrangian density in $\{t, r\}$ can be schematically  written
\ba
	 \mathcal{L}_{B} &=& \sum_{i=1}^{13} D_i(t,r,\ell) {\cal B}_{i}(h_{a_i},h_{b_i}),
	 	\label{OddLagr}
\ea
where the $D_i$ coefficients depend on the background and the total angular momentum $\ell$ but not $m$. ${\cal B}_i$ represents a bilinear operator  with at most one derivative  in
$t$ or $r$  on each of the fields $h_{a_i}, h_{b_i} \in \{ h_0, h_1, h_2 \}$. 
 Explicit expressions for these
terms are provided in Eqs.~(\ref{OddLagrRepeat}) and (\ref{eqn:Ds}).
For example ${\cal B}_{13} = h_0 \dot h_2$ and comes from the term $h^{\mu\nu;\alpha}
h_{\nu\alpha;\mu}$ of Eq.~(\ref{EinsteinHilbert}) with $\alpha = t$ and $\mu, \nu$ angular
coordinates or $\mu = t$ and $\alpha, \nu$ angular. Focusing on the former case, 
$h^{\mu\nu;\alpha}$ contains  $\dot h_2 Y^B_{\ell
m,ab}$ and $h_{\nu\alpha;\mu}$ contains $h_0 \nabla_a Y^B_{\ell m, b}$.  Since the covariant derivative on the
sphere $\nabla_a$ raises and lowers the spin weight, the orthogonality of 
angular integrals \eqref{eqn:angularidentities} produces the coupling.

It is clear from the number of terms in Eq.~(\ref{OddLagr}) and the explicit form \eqref{eqn:Ds} for their coefficients
that just extracting the expected spin 1 and 2 degrees of freedom is difficult and finding their characteristics even more so.
Unlike in general relativity, no further simplifications are possible since we cannot  eliminate
coupled modes  utilizing gauge freedom of the theory (see \S\ref{sec:GR}).

\subsection{$\ell = 1$}
\label{ssec:oddl1}

Since only $s \le \ell$ fields exist at a given $\ell$, the odd $\ell=1$ case is special in that the spin-$2$ field $h_2$ is
not present and we expect only one of the remaining fields $h_0, h_1$ to propagate after applying all the constraints.
Following our algorithm outlined in \S\ref{sec:RW} and detailed in 
\S\ref{sec:AlgorithmAppendix},
we first rewrite
the two EOMs into a first order system by introducing four additional
fields $h_{0t}, h_{0r}, h_{1t}$, and $h_{1r}$ corresponding to derivatives indicated by the second subscript
and add their definitions to the EOMs, e.g.\ 
\be 
	\dot h_0 - h_{0t} = 0, 
	\quad  h_0' - h_{0r}=0.
	\label{eqn:h0consistency}
\ee  
We then arrive at a set of six first order
differential equations which can be captured in the form \eqref{eqn:vecform}.

Instead of proceeding directly to the full 
Kronecker decomposition, we can first look for all $\Lmat_1^P$ overdetermined blocks by
noticing combinations of equations without temporal
or spatial derivatives and matching these together (see \S\ref{sec:AlgorithmAppendix}).
In fact just by inspection we know that the Kronecker structure of the equations contains at least two $\Lmat_1^P$ overdetermined blocks corresponding to
Eq.~(\ref{eqn:h0consistency}) and its $h_1$ counterpart.  In general, each $\Lmat_1^P$ block hides one constraint.
Here these  are the consistency relations
\ba 
	\dot h_{0r} - h_{0t}' &=& 0,  \nonumber\\
	\dot h_{1r} - h_{1t}' &=& 0,
\ea
which we add to the EOMs.

At this point we have eight EOMs for six field variables, leading to an $8\times 6$ matrix
pencil \eqref{eqn:vecform}.  The $\Lmat_1^P$ discovery process also identifies a  block associated with
\ba
\label{OddKroneckerExample}
	\dot \psi - \omega_1 &=& 0 ,\nonumber\\
	\psi ' - \omega_2 &=& 0 ,
\ea	
related to the overdetermined variable
\be
	\psi = h_{1t} - h_{0r} .
\ee	
In the formula above, $\omega_i$ are two linear combinations of the fields without any
derivatives; their particular form is not important. The block of equations
\eqref{OddKroneckerExample} contains a hidden constraint in a form of the first order
differential equation
\be
	\omega_1' - \dot \omega_2 = 0 .
\ee	
This equation is added to the investigated system, which is now described by a $9 \times 6$ pencil. 
Its Kronecker form now contains
\begin{equation}
\{  \Lmat_2, \Lmat_0^P,  3\times \Lmat^P_1 \} .
\end{equation}
In general, $\Lmat_0^P$ structures represent algebraic constraints that contain no derivative terms.
 Assuming that
 $\OurG_{12} \neq 0$, we can use the constraint to
integrate out $h_{1t}$ completely 
 and at the same time remove one of the nine EOMs which is  due to the redundancy
caused now by the removal of $h_{1t}$. 
This operation turns the underdetermined
$\Lmat_2$ block into a regular block $\Rmat_2$ 
and the Kronecker form into the
$8 \times 5$ system
\begin{equation}
\{3 \times  \Lmat^P_1, \Rmat_2\big(\tfrac{\OurG_{12}}{\OurG_{11}}\big) \}
\end{equation}
which contains no underdetermined blocks.   Furthermore it contains no hidden constraints
since we have extracted  one hidden relation from each of the three over determined $\Lmat_1^P$ blocks.

Using the third step of the algorithm, we identify from $\Rmat_2$
a single characteristic  of degeneracy 2 that is defined as integral curve
 $t(r)$ of
\be 
	\label{oddl1cha} \frac{dt}{dr} = - \frac{\OurG_{12}}{\OurG_{11}} .  
\ee
We interpret this as one physical degree of freedom (two pieces of initial data or 
phase space degrees of freedom) corresponding to the odd parity of the spin-1
mode of the massive graviton.

\begin{figure*}
\center
\includegraphics[width = 0.99\textwidth]{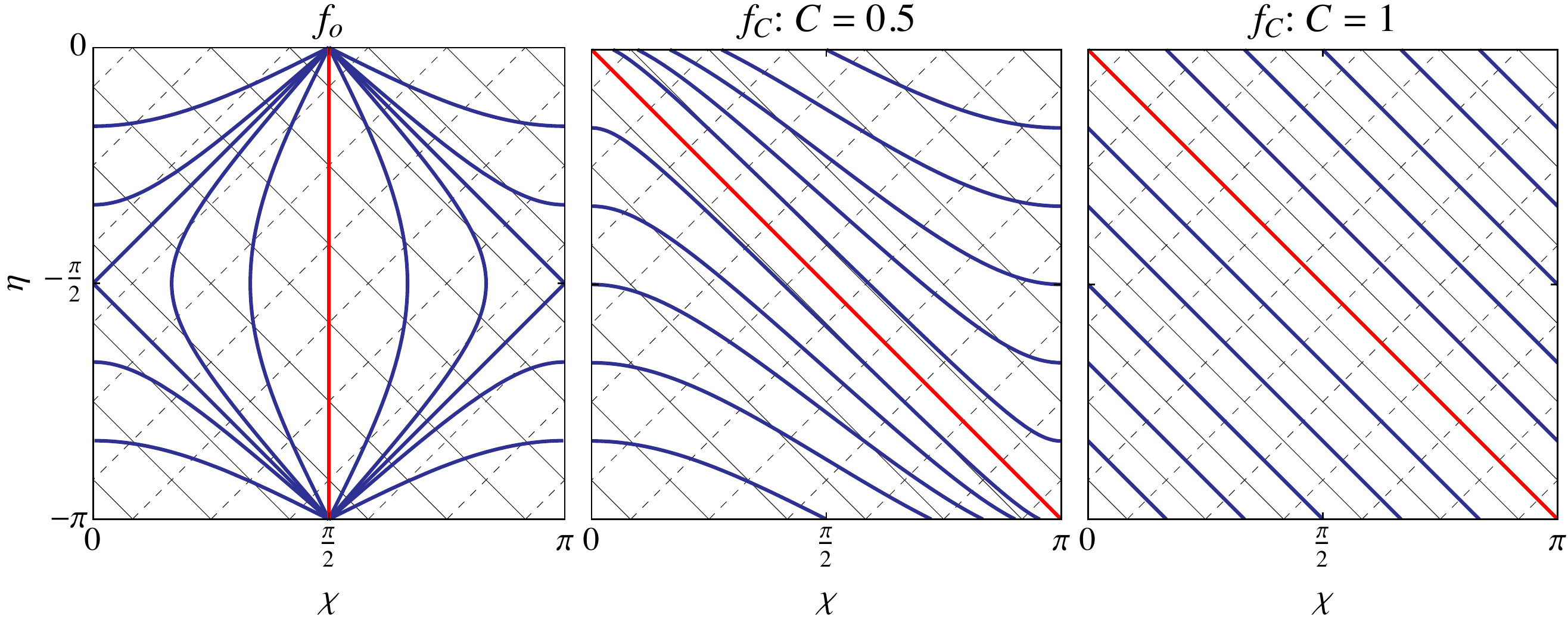}
\caption{Conformal diagrams of the characteristic curves for the $f_o$ background solution and two members of the $f_C$ family of solutions.  Thick blue lines correspond to the new spin-0 and
spin-1 modes introduced in dRGT whereas thin solid and dashed lines
represent the luminal characteristics of the spin-2 modes.   dRGT modes all share the same
repeated characteristics but come from  both hyperbolic and parabolic blocks.    Except in special
cases, all modes of the same parity and angular momentum are coupled by non-derivative
terms.  Thick red lines represent determinant singularities.
 }
\label{fig:characteristics}
\end{figure*}

Spacetime diagrams of the characteristic curves in the $(\eta, \chi)$ coordinates for the background
solutions \eqref{solMukohyama} and \eqref{solKoyama} for $\OurG_{\mu\nu}$
are plotted in
Figure~\ref{fig:characteristics} (thick blue lines).   Regions of the de Sitter space are
divided into separate copies of the background solutions at the determinant singularity (red thick lines).
The $f_{C=1}$ case is the unique background solution where the
characteristics are everywhere luminal; it can be shown \cite{Motloch:2015gta} that all
other background solutions show regions of spacelike
characteristic curves around the poles $\chi = 0, \pi$.

Although the curves themselves coincide with that for isotropic or even $\ell=0$ perturbations 
obtained previously in Ref~\cite{Motloch:2015gta}, as will be shown with the current method
in \S\ref{sec:Even}, the boundary value problem here is very different.  
For this odd $\ell=1$ mode, $\Rmat_2$ indicates that the associated degree of freedom is
parabolic not hyperbolic. In this sense the boundary problem is similar to the
heat equation where one specifies field data on a spacelike characteristic surface
and uses the EOMs to march forward in time given
spatial boundary conditions at the ends of characteristics. 
 Note though that parabolic characteristics need not be spacelike.  
For example in the $f_o$ solution, these characteristics are timelike in the inner diamond (Fig.~\ref{fig:characteristics}, left panel).

We shall see that in the even $\ell=0$ case, the system contains two hyperbolic phase space degrees of
freedom that propagate on the same curves.   In this case, the EOMs do not evolve
fields off the  characteristics and so field data must be specified
on a noncharacteristic surface.

Although the parabolic nature of the system is robust to field redefinitions, the field content of the overdetermined
and regular blocks is not. 
In particular, we can mix fields from
overdetermined blocks to regular blocks  since they are non-dynamical.    We can also mix variables between regular blocks
of the same characteristic, but not of different characteristics,
as detailed in \S\ref{ssec:KroneckerAmbiguity}.   For example, in the discussion
above, we are led to the assignment 
\be
	\vvec = \left( h_0, h_1,  \psi, h_{0r} {-} \tfrac{\OurG_{11}}{\OurG_{12}}h_{1r}, 
	{-}\tfrac{\OurG_{11}^2}{\OurG_{12}^2}h_{1r} \right)^T \!\!\!\! ,
\ee	
where the first three variables correspond to the 3 overdetermined blocks, leaving the last two as
the nominal ``propagating" degrees of freedom. 
However, an equally valid
representation with the same Kronecker structure is for example
\begin{equation}
	\vvec = \left( h_0, h_1,  \psi,
	 h_{0r} {\scriptstyle -} \tfrac{\OurG_{11} }{\OurG_{12}}h_{1r}
	{\scriptstyle -} \tfrac{b'}{b} h_0, 
\,
	{\scriptstyle -}\tfrac{\OurG_{11} }{\OurG_{12}} h_{0r}{\scriptstyle +} 
	 \psi\right)^T \!\!\!\!.
\end{equation}

More usefully, it is possible to show that
for a particular choice of $\Pmat, \Qmat$ the field variable $v_4$ corresponding to the first field 
in the parabolic block completely decouples from the remaining four fields, 
forming an autonomous equation 
\be
\label{Autonomous}
	-\frac{\OurG_{12}}{\OurG_{11}} \dot v_4 + v_4' +	C v_4 = 0
\ee
for some $C(t,r)$ which will not be given here. This is in full agreement with the
alternative analysis of \S\ref{ssec:oddal}, which also finds that one of the fields obeys
an autonomous equation~\eqref{eqq} with the same characteristic.
In this special case, explicit solutions for $v_4$ may be obtained by integrating data
from a non-characteristic curve as {\eqref{Autonomous}} is itself a decoupled hyperbolic equation.   Note that the EOM associated with 
$v_5$ still remains coupled to $\dot v_4$ so its initial or boundary data cannot be given independently
of this solution.

\begin{center}
\begin{table}
\caption{Characteristics and multiplicity of regular blocks.  }
\begin{tabular}{cccccc}
\hline\hline
$\, X \,$ & $\quad\ell\quad$ & $\Rmat_1\big( \frac{\OurG_{12}}{\OurG_{11}}\big)$ &  $\Rmat_2\big( \frac{\OurG_{12}}{\OurG_{11}}\big)$
&   $ \Rmat_1\big(\frac{a}{b}\big)$ & $\Rmat_1\big({\scriptstyle -}\frac{a}{b}\big)$ $\vphantom{\Big(}$ \\
\hline
$B$ & 0 &0 & 0 & 0 & 0 \\
$B$ & 1 & 0 & {1} & 0 & 0 \\
$B$ & $\ge 2$ & 0 & {1} & 1 & 1\\
$E$ & 0 &2  & 0 & 0 &0 \\
$E$ & 1 & 2 & 1 & 0 & 0 \\
$E$ & $\ge 2$ & 2 & 1 & 1 & 1 \\
\hline
\end{tabular}
\label{tab:Characteristics}
\end{table}
\end{center}

\subsection{$\ell \geq 2$}
\label{ssec:oddl2}

Although we have all three fields present at $\ell \geq 2$, the analysis is basically the same as for $\ell = 1$.
The difference is that we add three additional fields $h_2,h_{2t}, h_{2r}$, 
and four equations: the EOM for $h_2$, the definitions of
$h_{2t}, h_{2r}$ and the constraint associated with this new  $\Lmat_1^P$ block
\be
	\dot h_{2r} - h_{2t}' = 0.
\ee
The other structures in the system are the same as with $\ell=1$ treatment.  The $\Lmat^P_1$ block
related to $\psi$ reveals a $\Lmat^P_0$ block which allows us to integrate out $h_{1t}$.
With these constraints, the system is described by a $12\times 8$ Kronecker form
\begin{equation}
\{ 4\times  \Lmat^P_1, \Rmat_2\big(\tfrac{\OurG_{12}}{\OurG_{11}}\big),
\Rmat_1\big({\scriptstyle -}\tfrac{a}{b} \big),\Rmat_1\big(\tfrac{a}{b} \big) \} .
\end{equation}
The first regular block is parabolic and already present at $\ell = 1$; we associate it with the
odd parity spin-1 mode of the graviton. The other two regular blocks
are hyperbolic with characteristic curves 
\be 
\label{oddl2cha} \f{dt}{dr} = \pm \f{a}{b},
\ee 
which are luminal and directed radially inward or outward. These modes correspond to the
spin-2 graviton mode and necessarily contain the combination of fields  
\begin{equation}
\label{OddSpin2Fields}
h_{2r} \pm
\frac{a}{b} h_{2t}, \qquad (\mbox{hyperbolic, luminal}).
\end{equation}
Luminality of these curves is expected as this mode is inherited
from  general relativity which has the same kinetic structure as dRGT.

This association with the spin-2 mode \eqref{OddSpin2Fields} cannot be removed using the freedom in
performing the Kronecker decomposition, since fields in the
other regular blocks have different characteristics and the overdetermined blocks 
do not contain either $h_{2t}$ or $h_{2r}$.
Moreover there are non-derivative couplings of these modes to the other modes
through $\Cmat$ that also cannot in general be removed.

{\it Open solution.---}  
It is instructive to examine the case of the $f_o$ background solution in detail since its
homogeneity in open slicing permits a traditional scalar-vector-tensor (SVT) analysis
there \cite{Gumrukcuoglu:2011zh}.  In this slicing, the constant time surfaces coincide
with the $\OurG_{12}/\OurG_{11}$ characteristics.  In fact, our result seems paradoxical
in that SVT normal modes of the Laplace operator should fully decouple from each other in
linear theory.

To directly compare these results,  we reperform our analysis in open frame where $\OurG_{22} =
0$, a case that was excluded in our primary analysis but allowed by the technique itself
(see \S\ref{sec:eomchi}). This analysis therefore applies specifically to the open wedge
of de Sitter (upper right triangle of Fig.~\ref{fig:characteristics}, left panel) and
matches the domain investigated in Ref.~\cite{Gumrukcuoglu:2011zh}. We will skip
the details, as all proceeds similarly to our main analysis; as was argued before the
structure of the regular blocks is coordinate invariant and thus the same in the two analyses.

Using the freedom in the Kronecker form, we can in {this particular case} choose  $\vvec$ such that the fields of a parabolic block
$\Rmat_2$ and one of the $\Lmat_1^P$ blocks
completely decouple from the fields of the remaining five blocks. The EOMs which govern the
corresponding fields can then be solved independently.   In this sense the spin-1 vector mode does decouple
from the spin-2 tensor mode.    However these fields  then source those
in the remaining five blocks, in particular the spin-2 mode.
The resolution to the paradox is that solutions to the EOMs of the decoupled
blocks diverge at either origin or
the spatial infinity in ways that cannot be represented by the vector normal modes of the
Laplace operator.   
In other words, the usual SVT normal mode analysis sets these modes to zero by boundary
conditions, eliminating the source to the tensor modes.

This  one-way decoupling of the $\ell\ge 2$ odd parabolic block is not a general feature of the dRGT
self-accelerating solutions. More typically, the spin-1 and spin-2 variables mutually 
source each other and no simplification of the full system of
EOMs is possible.  Furthermore decoupling within the parabolic block that is possible for $\ell=1$
odd modes for all background solution (see Eqs.~\eqref{Autonomous} or \eqref{eqq}) is typically not
possible for $\ell \ge 2$ odd modes.

\subsection{Comparison with alternative analysis}
\label{ssec:oddlalt}

In Appendix~\ref{ssec:oddal}, we present an alternative analysis of the odd modes.
Introducing auxiliary fields, we  recast the odd Lagrangian as a second order system
for a new variable $\qa$ and $h_2$, with $h_0, h_1$ integrated out {for $\ell \ge 2$}.   
As such, there are no hidden constraints between the new variables and the system  can be
investigated by standard methods.

In particular, given the second order system EOMs, we perform a characteristic analysis by
searching for curves where discontinuities in the highest derivatives can occur.  This
analysis agrees on the spacetime trajectories and total multiplicities of the
characteristics.   For $\ell = 1$, integrating out $h_0$ leaves EOMs for $\qa$ and $h_1$ that
again confirm our main analysis.

These alternate analyses however fail to automatically find the distinction between parabolic
blocks and repeated hyperbolic blocks of the same characteristics which requires retention
of the first order derivative structure.
Moreover a drawback of modifying  the Lagrangian to resolve constraints
is that different types of constraints require different methods.   In fact as we shall
see in the next section, the even modes present such a complex constrained system that it
is unclear how to proceed at the Lagrangian level.   Our method provides an algorithmic
method of resolving hidden algebraic or differential constraints for arbitrarily complex systems at the EOM level.

On the other hand, the alternative analysis allows us to perform a Hamiltonian analysis of
the system, see \S\ref{ssec:oddl1Hamiltonian} for the particular case of $\ell=1$.

\section{Even ``E" modes}
\label{sec:Even}

In this section we finish our analysis of the perturbations around vacuum
cosmological solutions of dRGT by finding characteristic curves for the even parity or ``E" modes.  We
start with the special cases $\ell=0$ and $\ell=1$, the former of which was previously
investigated in \cite{Motloch:2015gta} by the \stucky method \cite{Wyman:2012iw}, and then proceed to the general
case with $\ell \geq 2$.

\subsection{Lagrangian}
\label{ssec:evenLag}

The even mode Lagrangian is of the form
\ba
	 \mathcal{L}_{E} &=& \sum_{i=1}^{55} E_i(t,r,\ell) {\cal E}_{i}(e_{a_i},e_{b_i}),
	 	\label{EvenLagr}
\ea
where  like the odd modes $E_i$ are the coefficients of  ${\cal E}_i$, a bilinear operator on
pairs of the seven $E$ modes 
\begin{equation}
e_{a_i}, e_{b_i} \in \{ H_0, H_1, H_2,\alpha, \beta, K, G \}
\end{equation}  
that contains at most one  derivative on each field with respect to $t$ or $r$.
Given that there are 55 distinct terms, they will not be presented explicitly here.

\subsection{$\ell = 0$}
\label{ssec:evenl0}

For the isotropic modes, only the spin 0 fields, $K$ and $H_i$ where $i =
0,1,2$, remain in the Lagrangian.  We then introduce eight first derivative fields $H_{it},
H_{ir}, K_t$ and $K_r$, along with their defining equations to have all EOMs manifestly first order. The defining equations naturally pair themselves into
$\Lmat_1^P$ blocks in the Kronecker decomposition just like for the
odd modes. We can therefore 
automatically add the hidden constraints corresponding to these four $\Lmat_1^P$ blocks;
these take form of consistency equations such as 
\be
	\dot H_{0r} - H_{0t}' = 0 .
\ee

After taking them into account, the Kronecker decomposition of the resulting $16 \times
12$  pencil reveals two additional $\Lmat^P_1$ blocks related to $K_r, K_t$. These two
blocks hide two additional equations, which are then included into the analysis. 
In the resulting 18 $\times$ 12 structure, there are two algebraic $\Lmat_0^P$ constraints.
We can use these two constraints to integrate out $K_r$ and $H_0$ completely and at the same time
remove two redundant equations. Four of the remaining 16 equations turn
into algebraic $\Lmat_0^P$ constraints, allowing us to remove $H_{1t}, H_{2t}, H_{0r}$ and $H_{0t}$
together with four equations. 
Of the remaining 12 EOMs two
are redundant,
allowing us to reduce the number of EOMs for the remaining six fields down to 10. 
The final system is
\be
	\left\{4\times \Lmat_1^P, 2\times\Rmat_1\big(\tfrac{\OurG_{12}}{\bar
	\chi_{11}}\big)\right\} .
\ee
All hidden constraints which can be derived from the four overdetermined blocks are
included and our analysis is thus finished.

The characteristic curves corresponding to the two regular blocks are  described by the 
same  
slope,
\be
\label{EvenIsotropicChar}
	\frac{dt}{dr} = - \frac{\OurG_{12}}{\OurG_{11}} ,
\ee
that was associated with  the spin-1 odd modes.
However,  unlike the odd modes  both regular blocks are hyperbolic,
allowing for solutions based on a set of initial data given on a non-characteristic surface.
Thanks to the large field transformation group associated with the Kronecker decomposition, one of the
hyperbolic blocks can be completely decoupled from the remainder of the blocks to describe
a field governed by an autonomous equation similar to Eq.~\eqref{Autonomous}.

This conclusion is in complete agreement with previous investigations of the isotropic modes
\cite{Wyman:2012iw,Motloch:2015gta}. The analyses there relied on solving
for \stucky and metric perturbations in isotropic gauge.   There  the special combination 
$\delta\Gamma=\delta g-x_0r\delta a$ obeys the same autonomous $\Rmat_1$ hyperbolic
form. The remaining isotropic \stucky perturbation $\delta f(t,r)$ also
satisfies a first order differential equation while the remaining metric fluctuations are governed by constraints. 
Characteristic curves corresponding to both propagating variables
are exactly \eqref{EvenIsotropicChar}. The isotropic gauge analysis is thus completely consistent
with our unitary gauge analysis.

As with the alternative $\ell=1$ odd mode analysis of \S\ref{ssec:oddal}, in the variables of the isotropic gauge analysis, a Hamiltonian
analysis is tractable.   Ref.\ \cite{Khosravi:2013axa} found that the Hamiltonian of this $\ell=0$ mode is unbounded from below. 
A Hamiltonian analysis for our unitary gauge system is intractable but from these two examples we can at least conclude
that there is no direct relation between the existence of an $\Rmat_1$  or $\Rmat_2$ block and an unbounded Hamiltonian.

\subsection{$\ell = 1$}
\label{ssec:evenl1}

The analysis is very similar to the $\ell=0$ one but with the addition of the spin-1 fields $\alpha, \beta$. 
As before, we introduce the 12 additional first derivative
fields and their defining equations. We again directly add the six consistency conditions
which correspond to the $\Lmat_1^P$ blocks related to the defining relations.

In this 24 $\times$ 18 system, we then find three new $\Lmat_1^P$ blocks 
whose hidden constraints then reveal three $\Lmat_0^P$ blocks.   The latter allow us to
integrate out $K_r, \alpha_r$ and $H_0$ which then further reveals  four algebraic constraints among the fields which
we use to remove $\beta_t, H_{2t}, H_{0t}$ and $H_{0r}$. As before, two of the remaining EOMs turn out to be  redundant and can be dropped.
The final Kronecker decomposition for the
$18 \times 11$ system becomes
\ba
	\left\{7\times \Lmat_1^P,2\times \Rmat_1\big(\tfrac{\OurG_{12}}{\OurG_{11}}\big)
	,\Rmat_2\big(\tfrac{\OurG_{12}}{\OurG_{11}}\big)
	\right\} .
\ea
There are no additional hidden constraints as all seven $\Lmat_1^P$ constraints are
already included.

On top of the blocks already present at $\ell = 0$  there is an $\Rmat_2$ block 
that we also found in the analysis of the $\ell=1$ odd modes. This agrees with our
interpretation that they together represent the two parity states of the spin-1 polarization of
the massive graviton.

\subsection{$\ell \geq 2$}
\label{ssec:evenl2}

For $\ell \ge 2$  there is an additional spin-2 field $G$ but the analysis is basically the same as for $\ell = 1$.
We first add three more fields $G, G_t, G_r$, two more defining conditions and the hidden
consistency relation
\be
	\dot G_r - G_t' = 0.
\ee
From there on the analysis follows exactly the same steps as the $\ell = 1$ analysis with the removal
of first $K_r, \alpha_r, H_0$ and  then $\beta_t, H_{2t},H_{0t}, H_{0r}$ with constraints 
and finally the dropping of two redundant EOMs.  The final Kronecker structure
\begin{equation}
	\big\{8\times \Lmat_1^P,
		2 \times \Rmat_1\big(\tfrac{\OurG_{12}}{\OurG_{11}}\big),
	\Rmat_2\big(\tfrac{\OurG_{12}}{\OurG_{11}}\big),
	\Rmat_1\big(\tfrac{a}{b}\big), \Rmat_1\big({\scriptstyle -}\tfrac{a}{b}\big)
	\big\} 
\end{equation}
contains the same regular components as $\ell=1$ corresponding to the spin-0 and spin-1 modes.
In addition there are two 
new regular blocks with luminal characteristic curves
\be
	\frac{dt}{dr}  = \pm \frac{a}{b}.
\ee
This completely mirrors the odd modes, where the difference between $\ell \ge 2$ and $\ell =
1$ also consists of a new luminally propagating mode \eqref{oddl2cha}. As we mentioned earlier,
luminality of the mode agrees with our expectations. The fields associated with these
luminal blocks are 
\begin{equation}
G_t \pm \frac{b}{a}G_r , \qquad (\mbox{hyperbolic, luminal}).
\end{equation}
and these blocks are thus unambiguously  related to the spin-2 field. Notice that in the $f_o$ solution
there are no spacelike curves that intersect the hyperbolic characteristics of all types whereas
in the $f_C$ solutions there are (see Fig.~\ref{fig:characteristics}).   Nonetheless the joint degrees of freedom in the $f_C$ case are not hyperbolic due to the parabolic blocks.

For the even modes it is possible to perform a similar decoupling analysis as we performed
for the odd modes. Again, for particular solutions such as $f_o$ it is possible to
completely decouple the parabolic block $\Rmat_2$ and one overdetermined block
$\Lmat_1^P$ from the rest. The parabolic modes 
diverge at either origin or spatial infinity and if not set to zero   \cite{Gumrukcuoglu:2011zh} will act as sources to the
luminal modes.    Again the parabolic decoupling does not happen for more general background solutions
and the spin-1 and spin-2 modes remain mutually coupled.

\section{Discussion}
\label{sec:discuss}

The dRGT theory of massive gravity presents interesting challenges for the study of metric 
perturbations around its vacuum self-accelerating backgrounds.   Although the background 
spacetime remains homogeneous and isotropic, the presence of a second metric can break translational invariance and invalidate the standard scalar-vector-tensor decomposition.   Furthermore,
the 10 metric variables are derivatively coupled and hide both differential and algebraic constraints
that permit just 5 independently propagating modes.  In this paper we have {
developed and employed} techniques to surmount these challenges.

Given the isotropy of the background and the parity invariance of the theory, 
we  use the Regge-Wheeler-Zerilli decomposition to decouple modes of different
angular momentum and parity.  The  equations of motion describe the propagation of the
coupled spin states of the massive graviton in the remaining radial dimension.

These equations of motion hide algebraic and differential constraints from the lapse, shift and
ghost-free construction of dRGT.  In Appendix~\ref{sec:AlgorithmAppendix} we develop
an algorithm to find its hidden constraints and characteristic curves.
Using this technique, we find the
new spin-0 as well as even and odd  spin-1 degrees of freedom all possess the same characteristic
curve
\be 
	\frac{dt}{dr} = - \frac{\OurG_{12}}{\OurG_{11}} ,
\ee
that depends only on the background solution not on angular momentum or parity.
These characteristics always run tangent to
determinant singularities.   On the other hand the two spin-2 degrees of freedom 
propagate on luminal characteristics.  This behavior of the tensor modes is
expected, because dRGT and general relativity share the same kinetic structure while they 
differ in their constraint structure.

Different spin states require different initial and boundary conditions.  
The scalar mode is hyperbolic and requires data on a surface that intersects all of
its characteristics.
For solutions like $f_o$ where there are no such surfaces  that are spacelike, the initial value problem
is ill-posed.   Moreover, the spin-2 modes are hyperbolic and a joint solution requires
a common surface that intersects all characteristics.

The two spin-1 modes propagate along the
same characteristic but each form a parabolic system, much like the heat equation. To
specify their evolution, we need to provide initial data on a characteristic surface and
supplement  it with two boundary conditions.

These conclusions agree with a previous analysis of the isotropic modes $\ell = 0$
using a different method \cite{Motloch:2015gta,Khosravi:2013axa} as well as a different analysis
of the $\ell=1$ odd modes presented in \S\ref{ssec:oddal}.   
In both of these cases, the Hamiltonian  is unbounded from below.

Finally within a given angular momentum and parity set,  various spin modes are still
coupled by non-derivative terms in ways that cannot be removed by field redefinitions
except in special cases.   The presence of coupled degrees of freedom that are both
hyperbolic and parabolic in nature and propagate on different characteristics implies
that the metric modes of dRGT cannot be evolved as a simple Cauchy problem.

\smallskip
Central to these analyses is the algorithmic method, presented with numerous examples in
Appendix~\ref{sec:AlgorithmAppendix}, to find hidden constraints and characteristic curves of an arbitrary
system of linear partial differential equations in 1+1 dimensions.   This method has 
a wide range of uses beyond the dRGT theory.

The logic of the method is first to rewrite all the equations of
motion into a first order system of differential equations. The Kronecker decomposition of the
resulting matrix pencil provides a systematic means of extracting hidden constraints and
identifying residual gauge freedom for the purpose of identifying the regular system that defines
the unique, consistent evolution of fields.  The generalized eigenvalues of the regular block
define characteristics. Their degeneracy and reality determines the hyperbolic,
parabolic and elliptic nature of the fields.

By identifying constraints hidden in derivatives of the
original equations of motion, these techniques should be useful in other systems where
the phase space degrees of freedom are reduced by constraints.

\acknowledgements{
We thank Austin Joyce, Teruaki Suyama,  Robert Wald and the organizers and participants of JGRG25 for useful discussions.  
This work was supported by U.S.~Dept.\ of Energy
contract DE-FG02-13ER41958. 
WH was additionally supported by
the Kavli Institute for Cosmological Physics at the University of
Chicago through grants NSF PHY-0114422 and NSF PHY-0551142 and NASA ATP NNX15AK22G.  
PM was additionally supported by grants NSF PHY-1125897 and NSF PHY-1412261 and thanks the Perimeter Institute for
Theoretical Physics where part of this work was performed. Research at Perimeter Institute is supported by
the Government of Canada through Industry Canada and by the Province
of Ontario through the Ministry of Economic Development \& Innovation.
}

\appendix

\section{Hidden constraints and characteristics}
\label{sec:AlgorithmAppendix}

We present here an algorithm which we use in revealing hidden algebraic and differential 
constraints as well as determining the characteristic curves. 
These are curves along which the equations of motion specify unique solutions for the
fields given their values at initial or boundary points.  
In principle, our technique can be used for
any set of linear partial differential algebraic equations (PDAEs) in 1+1 dimensions.

Analogously to the well-studied case of ordinary differential algebraic equations (DAEs), PDAEs
are partial differential equations where some of the equations represent constraints.
{Unlike  DAEs, where the constraints are purely algebraic, for PDAEs they can be algebraic in one of the dimensions and
differential in the other dimension.} Our method is thus well suited to analyze
perturbations around spherically symmetric solutions with complicated derivative couplings
and nonalgebraic constraints, such as dRGT. Despite its generality and applicability to a
wide range of problems, we are not aware of its publication before in the entire form
(see \cite{Martinson:1999aaa} for discussion of PDAEs with algebraic constraints, 
which provides pieces of our construction).

In \S\ref{sec:RW}, we provided an executive summary of the three step technique which we elaborate on
here with a discussion of fields propagating in regular blocks and illustrative examples.

\subsection{First order reduction}

The first step is to rewrite all equations of motion as first order partial
differential equations by introducing additional field variables corresponding to field
derivatives of the next to highest order. With each new field, we add into the
investigated system an equation which defines this variable.
For example, if any of the
equations of motion contains $\ddot  u$, we introduce an additional field $u_t$ and
supplement the equations of motion with 
\ba
\label{FirstConsistency}
	u_t &=& \dot u .
\ea
We can choose to introduce fields in a
symmetric fashion regardless of whether they are required for the initial reduction, i.e.~introduce both $u_t$ and
\ba
	u_r &=& u' .
\ea
This implies a ``hidden" consistency relation
\ba 
	u_t' -\dot u_r =0 ,
\ea
but we shall see that even if we do not add this consistency relation to the EOMs at the outset, it will be discovered
in the analysis.
We organize these equations into matrix form as
\begin{equation}
\Amat \dot \uvec+ \Bmat \uvec ' + \Cmat \uvec = 0.
\label{BasicEquation}
\end{equation}
This notation should not be confused with vectors and tensors in the spacetime or on the
2 sphere.

\subsection{Hidden constraints and Kronecker form}

The second step is to determine the independent propagating degrees of freedom
and form a set of equations specifying the unique evolution of a well formed system.
If after finding all algebraic and hidden constraints, the system remains underdetermined then
the system is ill-formed.

Suppose $\Amat$ is invertible.  Then we know that time evolution of the fields can
be specified by their values on spatial surfaces.   The generalization of this concept
for singular $\Amat$ 
is that if $\Amat + \lambda \Bmat$ is invertible for some choice of $\lambda$ then 
there is some suitable temporal coordinate, e.g. $t' = t+\lambda r$, where evolution is again 
defined.   In this case, $\Amat, \Bmat$ are said to form a regular pencil and
$\Amat + \lambda \Bmat$ is composed entirely by regular blocks.  
In such case, we can proceed directly to the next step to find the characteristics or the preferred temporal
coordinates associated with the different fields in the regular block.

If  $\Amat + \lambda \Bmat$ is singular for any $\lambda$, 
then the evolution naively looks ill-defined along any curve.
In general, a pencil can
be singular because the matrix system contains either over- or underdetermined blocks
or both.   For underdetermined blocks, solutions naively are not unique.   For
overdetermined blocks, solutions from arbitrary initial data can in principle be inconsistent.   Overdetermined blocks thus hide consistency relations which once exposed can convert underdetermined
blocks into regular blocks that yield a unique and consistent solution.

We therefore look for consistency relations associated with the overdetermined blocks
to augment the EOMs.    Before turning to the systematic approach, it is worthwhile
to discuss why constraints can be hidden in the overdetermined block.   First recall the case
of a first order ODE.   If some linear combination of
EOMs contains no time derivatives, then the remaining structure is either $0=0$  or  there is an algebraic
relationship between the fields.   In the former case, the system is trivially overconstrained 
indicating that an equation is redundant and can be removed.
In the latter case, one can solve for one of the
fields and eliminate it from the system or keep all the fields but add the 
derivative of the constraint equation to the EOMs. Note that the choice of which field to eliminate is somewhat
arbitrary and this will be related to a similar choice for the PDAE system.

In the case of a PDAE, the generalization is that there can also be equations that lack either temporal or spatial derivatives but not both.  In this case, the ``algebraic" constraint
is really differential in the other dimension.   It is then not straightforward to eliminate or
``integrate out" a field associated with the constraint.   On the other hand, derivatives of the constraint can still
add an independent EOM that evolves the constraint consistently.
This is similar 
to the presence of secondary constraints in a Hamiltonian analysis
or the differentiation index of a DAE.

The algorithmic way of proceeding is to  utilize the Kronecker decomposition
of the singular pencil  (see Appendix~\ref{sec:KroneckerAppendix}
for definitions and notation).   In terms of our PDAE, this amounts to choosing a particular
linear combination of fields $\vvec$ or field redefinition and linear combination of EOMs that 
exposes the regular, over and underdetermined blocks.  More specifically given 
appropriate invertible matrices $\Pmat, \Qmat$
\ba
	\vvec &=& \Qmat^{-1} \uvec , \nonumber \\
	\tilde \Amat &=& \Pmat\, \Amat\, \Qmat , \nonumber\\
	\tilde \Bmat &=& \Pmat\,  \Bmat\,  \Qmat , \nonumber
	\\
	\tilde \Cmat &=& \Pmat\, \Cmat\,  \Qmat + \Pmat\, \Amat\, \dot \Qmat + \Pmat\, \Bmat\, \Qmat' ,
\ea
we can rewrite Eq.~\eqref{BasicEquation} as
\be
	\tilde \Amat\, \dot \vvec + \tilde \Bmat\, \vvec ' + \tilde \Cmat \,\vvec = 0 ,
	\label{eqn:vveceqn}
\ee
with the matrix pencil $\tilde \Amat + \lambda \tilde \Bmat$ in the Kronecker form. Notice that 
$\Pmat$ describes particular linear combinations of the equations of motion while   $\Qmat$
performs linear field redefinitions. Because in general $\Qmat$ is a function of $\left\{ t, r \right\}$ the
linear combinations of the fields and equations corresponding to the individual Kronecker blocks 
depend on the position in the spacetime.

In Kronecker form, the matrix pencil is composed of blocks. 
Each $\Lmat_\mu^P$ block is
overdetermined --- there are $\mu$ fields and $\mu+1$ equations.
Conversely,
each $\Lmat_\mu$ block is  underdetermined  --- there are
$\mu + 1$ fields and only $\mu$ equations of
motion.

Each overdetermined block hides one constraint and if it is not already included in the EOMs,
we add it.   In the special case that $\mu=0$, corresponding to a row of zeros in the Kronecker form,
the constraint is algebraic as it would be for an ODE.  In that case, we either eliminate a field
or eliminate a redundant equation. If that resolves the only singular block, we repeat this step 
and form the new regular pencil.

The more novel cases are where there are $\mu \ge 1$ overdetermined blocks.   The case of
$\mu=1$ is instructive and is the only relevant one for dRGT given its particular second
order structure.
Here the equations in the block take the form
\ba
	\dot v_i +  c^j v_j &=& 0 ,\nonumber \\
	v_i' +  d^j v_j &=& 0 ,
\ea
for some coefficients $c^j, d^j$ where summation over repeated indices is implicit. We can subtract the time derivative
of the second equation from the spatial derivative of the first equation, again forming an
equation which is first order in the derivatives
\be
	 c^j v_j' - d^j \dot v_j + ({c^j}{}' - \dot d^j)v_j = 0 .
\ee
If this new EOM for the fields $v_j$ is not a linear combination of the existing ones, we add it to the list.   For systems where only $\mu=1$ overdetermined blocks
exist, these constraints can be found by inspection rather than by formal Kronecker decomposition.   They correspond to a linear combination of fields $v_i$ which obeys one equation with no time derivatives and another equation with
no spatial derivatives.   In practice to discover such combinations
it suffices to find all linear combinations of the
equations of motion which do not contain any spatial or temporal derivatives $s^i
{\rm EOM}_i$ and $t^i {\rm EOM}_i$ respectively. After we
discover all such combinations we can try to pair them in a way that
\be
	\frac{\partial}{\partial r}\( s^i {\rm EOM}_i\) - \frac{\partial}{\partial t}\( t^i {\rm EOM}_i\) 
\ee
contains no second order derivatives.  This is then a valid first order EOM that can be added to the system.

A similar but more involved procedure applies to overdetermined blocks with $\mu>1$.  Such structures
hide constraints between fields at higher order in derivatives and present opportunities to eliminate higher order
terms. In this case the combination of fields
$v_i$ which appears in the EOMs with no temporal derivatives is not the same as the
combination $v_j$ which appears in the EOMs with no spatial derivatives.   Instead the two are connected
by a derivative chain through the $\Lmat_\mu^P$ system of $\mu+1$ equations.  In this case by taking
$\mu-k+1$ spatial derivatives and $k-1$ temporal derivatives of the $k$th equation, one can construct a combination
with no $(\mu+1)$th order derivatives.   Like the $\mu=1$ system, this combination involves a constraint on the
system with $\mu$ derivatives.   The complication is that to cast the system in first order form,
auxiliary fields with $\mu-1$ derivatives must be introduced into the system.    Nonetheless, since this introduction 
amounts to fields with no extra freedom associated with them, the constraint represents a new equation which
if not already in the EOM system is added to them.    Since each $\Lmat_\mu^P$ block is a $(\mu+1) \times \mu$ matrix
system, once this constraint is found it exhausts the extra information in the overdetermined block.

After including the new information from all overdetermined blocks, we place the augmented system in its
final Kronecker form.   If there are no longer any underdetermined blocks, we proceed to the next step with
just the regular blocks.   Overdetermined blocks remain but  consistent evolution of their fields is now enforced in the
regular block.

If underdetermined blocks still remain, then the EOMs have no unique solution.
In physical systems this is often due to gauge freedom which has to be fixed.  In these cases, gauge fixing provides
new constraints.  If adding them to the EOMs and repeating this step 
produces a Kronecker decomposition with only regular and overdetermined blocks then we can again proceed to the next step (see \S\ref{sec:GR} for an example).

\subsection{Characteristics}

The regular blocks in the final Kronecker decomposition determine the characteristic curves of the system.  Consider the simplest regular block $\Rmat_1(\Omega)$. It describes the dynamics of
a single degree of freedom $v_i$, described by the equation
\be
	 v_i'  -\Omega \dot v_i + c^j v_j = 0 .
\ee
If $\Omega$ is real, this equation specifies derivative of $v_i$ along the direction
\be
\label{Characteristic}
	\frac{dt}{dr} = -\Omega ,
\ee
and so evolves the field along this direction. For this
reason the curve defined by Eq.~\eqref{Characteristic} is a characteristic curve.  
When the characteristic curve is aligned with
the time coordinate, formally we would have to set $\Omega = \infty$. Instead, the
Kronecker decomposition of such a regular block $\Rmat_1(\infty)$ is defined to be of the nilpotent form through Eq.~(\ref{eqn:Rnil}).

Regular blocks that are of dimension 1 and produce real characteristics are hyperbolic.  Data on a
non-characteristic surface that intersects these curves defines a unique solution by
integrating their values along the characteristic curves.  If this surface is spacelike
then the subsystem has a well-posed initial value  or Cauchy problem. If all hyperbolic blocks
share a common spacelike non-characteristic surface then their joint Cauchy problem is 
well-posed. If the $\Omega$ is
complex, then the block is elliptic and requires solution by relaxation from values on all
boundaries.

If the regular block is higher than dimension 1, then it is parabolic. The triangular form of $\Rmat_i$ produces 
a chain of equations 
\ba
\label{SymmetryEOM2}
	v_1' - \Omega \dot v_1 + \cdots &=& 0 ,\nonumber\\
	v_2' - \Omega \dot v_2 - \dot v_1 + \cdots &=& 0 ,\nonumber\\
	&\vdots& \nonumber\\
	v_{i}' - \Omega \dot v_{i} - \dot v_{i-1} + \cdots &=& 0 ,
	\label{eqn:chain}
\ea
where the dots in the equations stand for nonderivative terms.
The first equation determines the $v_1$ characteristic through $\Omega$ just like the hyperbolic counterpart.  The next variable
inherits the same characteristic but now supplies an evolution equation for $v_1$ off of the characteristic.  
This pattern continues through the chain.  Unlike the hyperbolic system
we can define  data for the fields on a given characteristic and march
forwards across characteristics.   On each characteristic, which is typically spacelike, information is communicated from one boundary to the other ``instantaneously" with respect to the marching direction.   Thus conditions must typically be specified
at both boundaries.  In this sense, a parabolic system is similar to an elliptic equation along the direction of the characteristic
while sharing the hyperbolic property of marching data but instead from one characteristic to another.   The nilpotent case
where $\Omega=\infty$ takes the same form but with time and space switched.
The system as a 
whole is hyperbolic if and only if all
regular blocks are hyperbolic.

Note that our analysis distinguishes between characteristics of two independent  regular blocks $\{ \Rmat_1(\Omega),\Rmat_1(\Omega)\}$ that just happen to be the
same and degenerate characteristics of a single $\Rmat_2(\Omega)$ regular block. 
For example, the former could represent two decoupled wave equations with luminal characteristics which is clearly
a hyperbolic system as a whole.
In the literature, based on the association with a single second order system repeated characteristics
themselves are often used as the definition of a parabolic system (see
e.g.~\cite{Hoffman:2001aaa}) but  this definition can not fully
distinguish all the possibilities.

\subsection{Field assignment}
\label{ssec:KroneckerAmbiguity}

While the Kronecker decomposition of the matrix pencil is uniquely determined, the
matrices $\Pmat, \Qmat$ themselves are not.  Since $\vvec=\Qmat^{-1} \uvec$ determines a specific linear combination of the original variables
$\uvec$ that can be associated with the various blocks, field assignment is not unique and so is not formally
a step in our technique.   On the other hand these transformations are useful for finding field combinations where
$\tilde \Cmat$ is as block diagonal as possible so that $\vvec$ is as decoupled as possible.
Formally there are further transformations $\tilde \Pmat$, $\tilde \Qmat$ that obey the group multiplication
property
\begin{equation}
 \tilde \Pmat \Pmat ( \Amat + \lambda  \Bmat) \Qmat \tilde \Qmat =
 \tilde \Pmat (\tilde \Amat + \lambda \tilde \Bmat) \tilde \Qmat = \tilde \Amat + \lambda \tilde \Bmat,
\end{equation}
or symmetry that leaves the Kronecker form invariant.

There are two useful transformations  $\tilde \Qmat$ that are worth noting.
First, $\tilde \Qmat$ can be chosen to add linear
combinations of fields in an overdetermined block to those in a regular block. For example, given a matrix pencil
in Kronecker form
\be
	\tilde \Amat + \lambda \tilde  \Bmat =  \{ \Lmat_1^P,\Rmat_1(\Omega) \} =  \begin{pmatrix}
		1 & 0\\
		\lambda & 0\\
		0 & \lambda - \Omega
		\end{pmatrix} ,
\ee
we have $\tilde \Pmat(\tilde \Amat + \lambda \tilde \Bmat)\tilde \Qmat = \tilde \Amat + \lambda \tilde \Bmat$ for all
\ba
	\tilde \Pmat &=&\begin{pmatrix}
		1 & 0 & 0\\
		0 & 1 & 0\\
		-C\Omega  & C & 1
	\end{pmatrix},\quad
		\tilde \Qmat = \begin{pmatrix}
		1 & 0\\
		-C & 1
	\end{pmatrix} ,
\ea
where $C(t,r)$ is an arbitrary real function. Given that the two columns correspond to $v_1,
v_2$ for $C = 0$ we now find that the regular block can correspond to any field combination
$\tilde v_2 = v_2 + C v_1$. In general variables in overdetermined blocks may always be added to regular blocks.

When there are two regular blocks with the same $\Omega$ or characteristic curve, we can
perform an additional transformation which keeps the Kronecker form invariant. Let us assume
the Kronecker decomposition reveals two regular blocks $\{ \Rmat_i(\Omega), \Rmat_j(\Omega)\}$.
Each block represents a derivative chain of the form (\ref{eqn:chain}).   For clarity, let us assign
the $v$ variables associated with the first as  $x_1,\ldots,x_i$ and to the second as $y_1,\ldots,y_j$.
Starting from $y_1$ and combining it with $x_{k+1}$, offset in its own chain by any  $\max(0,i-j) \le k  < i$, we can take 
sequential linear combinations
\ba
	\tilde x_{k+1} &=& x_{k+1} + C y_{1},\nonumber\\
	&\vdots& \nonumber\\
	\tilde x_{i} &= & x_{i} + C  y_{i-k},
\ea	
where again $C(t,r)$ is an arbitrary real function.   The evolution equations for $\tilde x_i$ are still of the
form \eqref{eqn:chain}. The Kronecker structure is thus unchanged by this operation despite the field
redefinition.  The corresponding
$\tilde \Pmat, \tilde \Qmat$ can easily be derived using these linear combinations.

Notice that we can have $i = j$ and $k=0$, where we take
linear combinations of the whole blocks, and also $x= y$, in which case we perform field
redefinitions within a single regular block by sequentially adding lower fields in the chain to higher fields.   Conversely, 
fields in regular blocks with different characteristics $\Omega_i\ne \Omega_j$ cannot in
general be mixed.

\subsection{Examples}

We now illustrate the procedure with several illustrative examples.  We begin with the canonical examples from second order
linear PDEs: the wave, heat and Laplace equations.  We then give an example of an underdetermined system: gravitational
waves in general relativity where the gauge is left unspecified.  Finally we provide  examples where hidden constraints
reduce the number of propagating degrees of freedom or the order of derivatives in a coupled set of EOMs by eliminating phase space degrees of freedom.

\subsubsection{Wave equation}
\label{sec:wave}

First, let us apply the above algorithm to the wave equation 
\be \ddot f- f'' = 0, \label{waveeq} \ee
whose Lagrangian is in the simplest form by
\be \mathcal{L} = \dot f^2 - f'^2. \label{Lwaves} \ee
Since complicated Lagrangians often hide simpler ones due to the presence of constraints,
 let us illustrate our algorithm with an alternate form
 \be
	\mathcal{L} = \dot f^2 - 4 f'h + 4 h^2 . \label{Lwaveu}
\ee
Obviously, in this case we can directly read off the EOM for $h$ 
\be
	h = \frac{f'}{2} ,
\ee
integrate $h$ out of the Lagrangian, and recover the standard form of \eqref{Lwaves}.
Our analysis below basically does the same, but through the algorithm above.

The equations of motion for the given Lagrangian \eqref{Lwaveu} read
\ba
	\ddot f - 2 h' &=& 0 , \nonumber\\
	-f' + 2 h &=& 0 .
\ea
We can reduce this system to first order form by 
introducing $f_t, f_r$ through
\ba
	 \dot f - f_t &=& 0 , \nonumber \\
	 f' - f_r &=& 0 ,
\ea
which we append to the EOMs written in terms of these fields
\ba
	 \dot f_t - 2h' &=& 0 , \nonumber\\
	 -f_r + 2 h &=& 0.
\ea
The matrix pencil for the fields $\uvec = (f,f_t,f_r,h)^T$ is
\begin{equation}
\Amat+\lambda\Bmat=\left(
\begin{array}{c c c c}
1 & 0 & 0 &0 \\
\lambda & 0 &0 &0 \\
0 & 1 &0 &-2\lambda \\
0 & 0 & 0 &0 \\
\end{array}
\right).
\end{equation}
The presence of a row of zeros indicates an $\Lmat_0^P$ overdetermined structure and hence a constraint.  
In this case it is a simple algebraic constraint $h=f_r/2$, consistent with our
earlier discussion. Eliminating $h$ with the constraint
we are left with
\ba
	{\rm EOM}_1&:&\ \dot f - f_t = 0 , \nonumber\\
	{\rm EOM}_2&:&\ f' - f_r = 0 , \nonumber \\
	{\rm EOM}_3&:&\ \dot f_t -  f_r' = 0 .
\ea
If we started with the usual form of the wave equation~\eqref{waveeq},
we would arrive directly to this set of equations after step 1.

Now for the field vector $\uvec = (f,f_t,f_r)^T$ the matrix pencil is
\begin{equation}
\Amat+\lambda\Bmat=\left(
\begin{array}{c c c }
1 & 0 &0  \\
\lambda & 0 &0  \\
0 & 1 &- \lambda \\
\end{array}
\right).
\end{equation}
This is a singular pencil representing the fact that there is no explicit evolution equation for $f_r$.  
However in addition to the  one underdetermined block represented by the third row there is one overdetermined block specified by the first column.

A simple rearrangement of rows, vectors and sign conventions would place
this in Kronecker form with one $\Lmat_1$ and one $\Lmat_1^P$ block,
but is not necessary to see that there is a hidden constraint.
There is only one linear combination of these EOMs which has no spatial derivatives, EOM$_2$ itself
and one which has no temporal derivatives, EOM$_3$ itself.
Matching these two together is straightforward and is accomplished by differencing the complementary derivatives,
generating a consistency constraint
\be
	{\rm EOM}_4:\ \partial_r ({\rm EOM}_{1}  )  - \partial_t ({\rm EOM}_2) = \dot f_r - f_t' = 0.
\ee
Given that there is only one $\Lmat_1^P$ block the addition of this independent equation completes the system
and supplies the missing evolution equation for $f_r$.

Adding the constraint to the EOMs, we now have the pencil
\begin{equation}
\Amat+\lambda\Bmat=\left(
\begin{array}{c c c }
1 & 0 &0  \\
\lambda & 0 &0  \\
0 & 1 &-\lambda \\
0 &-\lambda & 1 \\
\end{array}
\right).
\end{equation}
This pencil has the original $\Lmat_1^P$ block in the first column but instead of an underdetermined $\Lmat_1$ block we now have a $2 \times 2$ block that contains only regular pieces.    With
\ba
	\Pmat &=& \begin{pmatrix}
		1 & 0 & 0 & 0\\
		0 & 1 & 0 & 0\\
		0 & 0 & -\frac{1}{2} & -\frac{1}{2}\\
		0 & 0 & -\frac{1}{2} & \frac{1}{2}
	\end{pmatrix} , \quad
	\Qmat = \begin{pmatrix}
		1 & 0 & 0\\
		0 & 1 & -1\\
		0 & 1 & 1
	\end{pmatrix} ,
\ea
we can put the pencil into its 
Kronecker form 
\begin{eqnarray}
	\tilde \Amat + \lambda \tilde \Bmat &=& \{ \Lmat_1^P, \Rmat_1(1),\Rmat_1(-1) \}\nonumber\\
	&=&
	 \begin{pmatrix}
		1 & 0 & 0\\
		\lambda & 0 & 0\\
		0 & \lambda- 1& 0\\
		0 & 0 & \lambda + 1
	\end{pmatrix} .
\end{eqnarray}

In agreement with the expectations, the regular blocks possess two luminal characteristics
\be
	\frac{dt}{dr} = \pm 1 ,
\ee
with field content $\vvec = [f,(f_r + f_t)/2,(f_r-f_t)/2]^T$. Each block is hyperbolic
in the sense that we can propagate one boundary condition along the characteristic curve.  

In the overdetermined block we have equations for $\dot f$ and $f'$ that can be integrated
self-consistently on any curve given $f_r\pm f_t$.  In fact, the appearance of the fields
in the regular block is not unique.  Since we have implicitly integrated out $f$, we could
add an arbitrary mixture of it back into the dynamical fields $f_r \pm f_t + c_\pm f$
which mathematically does not change the Kronecker structure or the
characteristics. This illustrates the fact that $\Pmat$, $\Qmat$, and $\vvec$ are not
unique even though counting of the degrees of freedom and the identification of their
characteristics is.

\subsubsection{Heat equation}

As the next example, consider the heat equation
\be
	\dot f - f'' = 0 ,
\ee
where $f$ is usually associated with temperature.
Step 1 is the reduction to a first order system.   We can either choose to just introduce $f_r=f'$ or
also introduce $f_t=\dot f$ to obtain a symmetric set.

Let us start with the first case.
Here the field vector is $\uvec = ( f, f_r)^T$, the EOMs are 
\begin{eqnarray}
{\rm EOM}_1 &:&\ f' - f_r = 0 ,\nonumber \\
{\rm EOM}_2 &:&\ f_r' - \dot f  = 0 ,
\end{eqnarray}
and the corresponding matrix pencil is already in Kronecker form
\begin{equation}
\Amat+ \lambda\Bmat =  \Rmat_2(0)= \left(
\begin{array}{c c }
\lambda  & 0 \\
-1  & \lambda   \\
\end{array}
\right),
\end{equation}  
 with a single characteristic curve
\be
	\frac{dt}{dr} = 0 .
\ee
The repeated characteristics in the block also represent the well-known fact that the heat 
equation is parabolic. Characteristics are constant time slices and instead of defining initial conditions
on a non-characteristic surface, one specifies them on an initial time slice.   The second equation, which is the original 
EOM,  then propagates
this information forward in time.    To fully define the system, we also require two spatial boundary conditions since
information propagates instantaneously across the time slice.

Now consider the second case where we introduced $f_t$ as a second auxiliary field.  In that 
case the EOMs are
\ba
	{\rm EOM}_1&:&\ \dot f - f_t = 0 , \nonumber\\
	{\rm EOM}_2&:&\ f' - f_r = 0 , \nonumber\\
	{\rm EOM}_3&:&\ f_t - f_r' = 0 .
\ea
With the field vector $\uvec = (f, f_t, f_r)^T$,
 the matrix pencil is
\begin{equation}
\Amat+\lambda\Bmat=\left(
\begin{array}{c c c }
1  & 0 & 0\\
\lambda & 0 & 0   \\
0 & 0 & -\lambda \\
\end{array}
\right).
\end{equation}
This corresponds to an overdetermined block in the first column $\Lmat_1^P$, an underdetermined $\Lmat_0$ block in the
second column and a regular block in the third.

Again the overdetermined block hides the same consistency constraint as the wave equation
\be
	{\rm EOM_4}:\  f_t' - \dot f_r = 0
\ee
and completes the underdetermined block to a larger regular block
\begin{equation}
\Amat+\lambda\Bmat=\left(
\begin{array}{c c c }
1  & 0 & 0\\
\lambda & 0 & 0   \\
0 & 0 & -\lambda \\
0 & \lambda & -1
\end{array}
\right).
\end{equation}
The Kronecker form for this {pencil} is achieved by choosing 
\ba
	\Pmat &=& \begin{pmatrix}
		1& 0 & 0 & 0\\
		0 & 1 & 0 & 0\\
		0 & 0 & -1 & 0\\
		0 & 0 & 0 & 1
	\end{pmatrix} , \quad
	\Qmat = \begin{pmatrix}
		1 & 0 & 0\\
		0 & 0 & 1\\
		0 & 1 & 0
	\end{pmatrix} ,
\ea
and reads
\ba
	\tilde \Amat + \lambda \tilde \Bmat = \{\Lmat_1^P,\Rmat_2(0)\}= \begin{pmatrix}
		1 & 0 & 0\\
		\lambda & 0 & 0\\
		0 & \lambda & 0 \\
		0 & -1 & \lambda
	\end{pmatrix} .
\ea
In this method we recover the same regular block and characteristic curves
as before but now associated with  $(f_r,f_t)$.

\subsubsection{Laplace equation}

The Laplace equation 
\be
	\ddot f + f'' = 0 ,
\ee
is the canonical example of a system where there are no real characteristics in the regular block.
For consistency of our notation we keep labeling the coordinates $(t, r)$ though as we shall see
such a case does not have an initial value formulation and hence physically is associated with
problems with two spatial dimensions.

Since the only difference with the wave equation is a change in sign of $f''$, we skip directly to step 2
with the hidden constraint added
\ba
	{\rm EOM}_1&:&\ \dot f - f_t = 0 ,\nonumber \\
	{\rm EOM}_2&:&\ f' - f_r = 0 ,\nonumber \\
	{\rm EOM}_3&:&\ \dot f_t + f_r' = 0 ,\nonumber \\
	{\rm EOM}_4&:&\ f_t' - \dot f_r = 0 ,
\ea
which has the pencil for $\uvec = (f,f_t,f_r)^T$
\ba
	\Amat + \lambda \Bmat &=& \begin{pmatrix}
		1 & 0 & 0\\
		\lambda & 0 & 0\\
		0 & 1 & \lambda\\
		0 & \lambda & -1
	\end{pmatrix} .
\ea

In this case, we explicitly show how to construct $\Pmat$ and $\Qmat$ to highlight
how a set of real but coupled first order PDEs can lack real characteristics.
Since the first two rows are already in the correct form, we can  focus our
attention to the $2 \times 2$ lower right subblock
\be
	\Amat_2+\lambda\Bmat_2=	\begin{pmatrix}
		1 & 0\\
		0 & -1
	\end{pmatrix} +
			\lambda \begin{pmatrix}
		0 & 1 \\
		1 & 0
	\end{pmatrix} .
\ee
The associated Weierstrass form has the identity $\Imat_2$ for the $\tilde \Bmat_2$ matrix  so we switch
the rows by multiplying on the left with
\be
	\Smat_2 = \begin{pmatrix}
		0 & 1\\
		1 & 0
	\end{pmatrix} .
\ee
Since
\begin{equation}
\Smat_2  	\Amat_2 	 = 
	\begin{pmatrix}
		0 & -1\\
		1 & 0
	\end{pmatrix} 
	\label{eqn:RA}
\end{equation}
is diagonalizable,  it can be placed into Jordan form with its eigenvectors.  We can immediately see that the
eigenvalues of Eq.~(\ref{eqn:RA}) are imaginary and the eigenvector matrix $\Qmat_2$ is complex.  Explicitly the
$2\times 2$
block comes into canonical form with
\begin{eqnarray}
\tilde \Amat_2&=&\Qmat_2^{-1} \Smat_2 \Amat_2 \Qmat_2 = \Pmat_2  \Amat_2 \Qmat_2  , \nonumber\\
\tilde \Bmat_2 &=&\Qmat_2^{-1} \Smat_2 \Bmat_2 \Qmat_2 = \Qmat_2^{-1} \Imat_2 \Qmat_2 = \Imat_2,
\end{eqnarray}    
so that 
putting the blocks together, we have
\begin{equation}
	\Pmat = \begin{pmatrix}
		1& 0 & 0 & 0\\
		0 & 1 & 0 & 0\\
		0 & 0 & 1 & i\\
		0 & 0 & 1 & -i
	\end{pmatrix} , \quad
	\Qmat = \left( \begin{array}{rrr}
		1 & 0 & 0 \vphantom{\Big[}\\
		0 & -\tfrac{i}{2} & \tfrac{i}{2} \vphantom{\Big[} \\
		0 & \tfrac{1}{2} & \tfrac{1}{2} \vphantom{\Big[} 
	\end{array} \right),
\end{equation}
and
\begin{eqnarray}
\tilde\Amat+\lambda\tilde\Bmat&=& \{ \Lmat_1^P, \Rmat_1(i), \Rmat_1(-i) \} \nonumber\\ &=& \left(
\begin{array}{c c c }
1  & 0 & 0\\
\lambda & 0 & 0   \\
0 & \lambda-i &0  \\
0 & 0 & \lambda+i
\end{array}
\right).
\end{eqnarray}

Notice the absence of real characteristic curves, which is a well-known feature of the Laplace equation. There are no preferred paths of information propagation and so 
solution at each spacetime point influences the solution at all other points.  Thus the Laplace equation cannot
be solved by integrating initial data along characteristics as information from all boundaries determines the solution.  
An attempt to solve the system as a Cauchy problem is ill-posed since the normal modes grow exponentially;
typical initial data will blow up at the future time infinity unless boundary conditions are enforced there.

\subsubsection{GR and gauge freedom}
\label{sec:GR}

Next we investigate gravitational waves in general relativity  around a Minkowski
background in  spherical coordinates
\be
\label{eqn:Mink}
\dd s^2 = -\dd t^2  + \left(\dd r^2 + r^2 \dd\Omega_2^2\right) .
\ee
The expansion of the Einstein-Hilbert action 
is given by \eqref{EinsteinHilbert}. 
Therefore our RWZ analysis for dRGT also applies to this case if we set 
$\Lnorm=\Leff=0$ and $a=b=1$.  In particular Eq.~\eqref{OddLagr} for the odd modes reduces to 
\begin{eqnarray}
\label{MinkowskiLagr}
\mathcal{L}_{B}^{\rm GR} &=&
\frac{1}{4} ( \dot h_1 - h_0' )^2 
+\frac{1}{r} h_0\dot h_1+\frac{ \ell (\ell+1)}{4r^2} h_0^2
\\
&&
-\frac{ (\ell-1)(\ell+2)}{4r^2} h_1^2+\frac{1}{16r^2} \mk{ \dot h_2^2 - h_2'^2 + \f{2}{r^2} h_2^2 }
\nonumber\\
&&+\frac{ \sqrt{(\ell-1) (\ell+2)} }{4r^2} \mk{ h_0 \dot h_2 - h_1 h_2' + \f{2}{r} h_1 h_2 }.
\nonumber
\end{eqnarray}

The three equations of motion can be written as first order differential equations if we
introduce six additional fields $h_{it}, h_{ir}, i \in \{0, 1, 2\}$, and their defining
equations
\ba
	\dot h_i - h_{it} &=& 0 , \nonumber\\
	h_i ' - h_{ir} &=& 0 .
\ea
Each of these is an $\Lmat^P_1$ overdetermined block and they together contain three hidden
 consistency equations
\be
	\dot h_{ir} - h_{it}' = 0 .
\ee
There is  an additional $\Lmat^P_1$ block in the Kronecker
decomposition, however the additional constraint is a tautology.    We have thus incorporated all of the 
additional constraints.   

The  Kronecker decomposition then reads
\be
	\tilde \Amat + \lambda \tilde \Bmat =\{\Lmat_2, 4\times \Lmat_1^P, \Rmat_1(1), \Rmat_1(-1)\} .
\ee
The presence of the underdetermined $\Lmat_2$ block shows the equations of motion are not
sufficient to determine uniquely the evolution of all  fields.
This is
not surprising as the general relativity Lagrangian \eqref{MinkowskiLagr} is
diffeomorphism invariant and contains a gauge symmetry
\ba
	h_0(t,r) &\rightarrow& h_0(t,r) + \dot \Lambda(t,r) , \nonumber\\
	h_1(t,r) &\rightarrow& h_1(t,r) + \Lambda'(t,r) - \frac2{r}
	\Lambda(t,r) , \nonumber\\
	h_2(t,r) &\rightarrow& h_2(t,r) + 2\Lambda(t,r) , 
\ea
where $\Lambda(t,r)$ is an arbitrary function. Because in our analysis we
did not fix this gauge freedom, the redundant modes appear in the final Kronecker
decomposition.

The next step is  to remove this gauge freedom.   As an example, we choose $h_2 = 0$. 
This gauge constraint when added to the system
{as an EOM} is formally an $\Lmat_0^P$ overdetermined block. Following our algorithm
and eliminating $h_2$ from the system reveals two additional algebraic constraints
\be
	h_{2t} = h_{2r} = 0,
\ee
coming from one $\Lmat_1^P$ block which turned into two $\Lmat_0^P$ blocks upon
integrating $h_2$ out. Eliminating these constraints is the same as erasing
$h_2$ from the original higher order EOM system at the outset. Notice that
gauge fixing after obtaining the EOMs still retains an equation of motion associated with
$h_2$; we comment on this subtlety below.
After eliminating these three variables, 
the Kronecker system is 
\be
	\tilde \Amat + \lambda \tilde \Bmat =\{\Lmat_2, 2\times\Lmat_0^P,3\times \Lmat_1^P\} .
	\label{eqn:GRintermediate}
\ee
The $\Lmat_0^P$ blocks represent algebraic constraints that can be used to complete the
underdetermined block to a regular block.  In the
 previous Kronecker decomposition these equations correspond to the $\Rmat_1(\pm 1)$
 blocks which now both turn into 
the same algebraic constraint $\Lmat_0^P$
\be
	h_{0t} - h_{1r} = 0 ,
\label{GaugeFixedEOM}
\ee
which means one of them is  redundant.
Elimination of $h_{0t}$ and the redundancy turns a previously underdetermined $\Lmat_2$ block into two regular $1\times 1$ blocks. At this point, no underdetermined
blocks remain, all the information in the overdetermined blocks is extracted and the analysis
is finished; the final decomposition reads
\be
	\tilde \Amat + \lambda \tilde \Bmat = \{3\times \Lmat_1^P,
	\Rmat_1(1), \Rmat_1(-1)\} .
\ee
The two regular blocks have luminal characteristics, as expected.

Finally, this example also illustrates a subtlety about gauge fixing.   Gauge fixing can always  be safely performed at the equations of motion level. 
Notice though that if the $h_2=0$ gauge were fixed directly at the Lagrangian level, we would never
vary with respect to it and would lose an equation of motion.  
Gauge fixing directly in the Lagrangian should only be performed if the equation of motion that is 
lost is redundant.   In  cases where it is not,
the system of  remaining
EOMs is incomplete and does not fully describe the physical system (see \S III\,C of
\cite{Motloch:2014nwa} for examples).

In the case considered here, we arrive at correct answer even when
we fix the gauge through setting $h_2 \rightarrow 0$ in the Lagrangian. This is because 
the information contained in the gauge-fixed $h_2$ EOM is
exactly  Eq.~\eqref{GaugeFixedEOM}. The $h_2$ EOM  is in fact responsible for the redundancy of the identical $\Lmat_0^P$ blocks in
Eq.~(\ref{eqn:GRintermediate}).  More generally, one can set a field to zero by using
gauge freedom if its gauge transformation does not
involve derivatives of the gauge function.\footnote{We thank Teruaki Suyama for discussion
on this point.} For this reason, the gauge fixing to unitary gauge in the dRGT quadratic Lagrangian using
Eq.~(\ref{eqn:unitary}) is a valid procedure which does not lose information in the \stucky EOMs.

\subsubsection{Propagating vs derivatively-constrained fields}
\label{sec:proderi}

A general quadratic Lagrangian for two fields with maximally two derivatives
typically propagates two degrees of freedom. However, as we show with the following example,
this counting can be mistaken due to hidden derivative constraints. The Lagrangian for dRGT described in
the main text is a more advanced case of the same phenomenon.

Let us investigate the Lagrangian
\be
\label{SpecialLagr}
	\mathcal{L} = \(\Hv_0' + \Hv_1' + \dot \Hv_0\)^2 - 2 \Hv_0^2 + 8 \Hv_1^2 .
\ee
The two equations of motion contain second derivatives, therefore we introduce four
additional fields $\Hv_{0r}, \Hv_{0t}, \Hv_{1r}, \Hv_{1t}$, where as before subscript determines
which derivative is taken, as well as the definitional EOMs
\ba
\label{ExampleE1}
	{\rm EOM}_1&:& \dot \Hv_{0} -\Hv_{0t}= 0,\nonumber\\
	{\rm EOM}_2&:& \Hv_{0}' -\Hv_{0r}= 0,\nonumber\\
	{\rm EOM}_3&:& \dot \Hv_{1} -\Hv_{1t}= 0,\nonumber\\
	{\rm EOM}_4&:& \Hv_{1}' -\Hv_{1r}= 0,
\ea
associated with them.
As usual these definitions provide $2\times \Lmat_1^P$ blocks that hide the consistency
constraints
\ba
\label{ExampleE2}
	{\rm EOM}_5&:& \Hv_{0t}'-\dot \Hv_{0r} = 0,\nonumber\\
	{\rm EOM}_6&:& \Hv_{1t}'-\dot \Hv_{1r} = 0.
\ea
Finally the original equations of motion in the first order variables can be written as
\begin{eqnarray} 
\label{ExampleE3}
	{\rm EOM}_7&:& 2\Hv_0 + \Hv_{0r}' + \Hv_{1r}'+2 \dot \Hv_{0r} + \dot \Hv_{0t} + \dot \Hv_{1r} = 0,\nonumber\\
	{\rm EOM}_8&:& 8\Hv_1 - \Hv_{0r}'-\Hv_{1r}'-\dot \Hv_{0r} = 0 .
\end{eqnarray}
This system of equations has the structure
\begin{equation}
\tilde \Amat + \lambda \tilde \Bmat = 
\{ \Lmat_2, 3\times \Lmat_1^P \}
\end{equation}
and so hides an additional $\Lmat_1^P$  constraint associated with 
\begin{equation}
Q=\Hv_{0r}+\Hv_{1r}+\Hv_{0t},
\label{eqn:EOMQ}
\end{equation}
exactly the combination that appears in the
term in brackets of Eq.~\eqref{SpecialLagr}.
Equating its mixed derivatives gives 
\ba
\label{ExampleE4}
	&&\partial_t\({\rm EOM}_8 - {\rm EOM}_5\) + \partial_r\({\rm EOM}_7 + {\rm EOM}_8\) =
	\nonumber\\
	&&\phantom{\partial_t({\rm EOM}_2} 2\(\Hv_{0r} + 4 \Hv_{1r} + 4 \Hv_{1t}\) =0.
\ea
This algebraic constraint allows us to integrate out $\Hv_{1t}$ and convert the $\Lmat_2$ block
into regular blocks.  The final Kronecker decomposition of the 8 equations and 5 variables reads
\be
\label{SpecialKronecker}
	\tilde \Amat + \lambda \tilde \Bmat = \{3\times \Lmat_1^P,
	\Rmat_1(-2), \Rmat_1\big({-}\tfrac{2}{3}\big)\} .
\ee
Notice that the system contains two regular blocks indicating just a single
propagating degree of freedom. Two initial conditions on a joint noncharacteristic
surface supplies sufficient information to determine uniquely the evolution of the
system, despite the form of the Lagrangian \eqref{SpecialLagr} which contains two fields
with second derivatives.    The specific construction above leads to the field combinations
in the various blocks
\begin{equation}
{\bf v} = \{q_0,q_1,Q,\tfrac{3}{4} q_{0r}-\tfrac{3}{2} q_{1r} , -\tfrac{3}{4} q_{0r} -\tfrac{3}{2} q_{1 r}  \}^T.
\end{equation}
It is useful to recall for the comparison that follows that the full EOMs can be constructed
as $\tilde\Amat \dot {\bf v} + \tilde \Bmat {\bf v}' + \tilde \Cmat {\bf v}=0$, where here
\be
	\tilde \Cmat =
	\begin{pmatrix}
 0 & 0 & -1 &\frac{1}{3}  &-1 \\
 0 & 0 & 0 & -\frac{2}{3} & \frac{2}{3} \\
 0 & 0 & 0 & -\frac{1}{6} & -\frac{1}{2} \\
 0 & 0 & 0 &\frac{1}{3} & \frac{1}{3} \\
2& 8& 0 & 0 & 0 \\
 0 & -8 & 0 & 0 & 0 \\
 0&-12 & 0 & 0 & 0 \\
 0 &4 & 0 & 0 & 0 
	\end{pmatrix} .
	\label{eqn:Cexample}
\ee

This example also illustrates the alternative analysis in Appendix~\ref{ssec:oddal}  in a simpler setting.  The Lagrangian \eqref{SpecialLagr} is equivalent to 
\be \label{SpecialLagrQ} \mathcal{L} = -Q^2 + 2Q \(\Hv_0' + \Hv_1' + \dot \Hv_0\) - 2 \Hv_0^2 + 8 \Hv_1^2, \ee
as we can recover \eqref{SpecialLagr} by plugging in the EOM for $Q$, i.e.~Eq.~(\ref{eqn:EOMQ}).  After integration by parts, the $\Hv_0$ and $\Hv_1$ EOMs become constraints
\begin{eqnarray}
\Hv_0 &=& -\frac{1}{2} (Q' + \dot Q), \nonumber\\
\Hv_1 &=& \frac{1}{8} Q',
\label{eqn:integrateout}
\end{eqnarray}
which are themselves  linear combinations of the overdetermined EOMs for $Q$
associated with the $\Lmat_1^P$ block given by the 5th and 6th rows of Eq.~(\ref{eqn:Cexample}).
We can then rewrite \eqref{SpecialLagrQ} as 
\be \mathcal{L} = \f{1}{2}\dot Q^2 + \dot Q Q' + \f{3}{8} Q'^2 - Q^2 ,\ee
which gives an EOM
\be 
\label{ExampleE5}
\ddot Q + 2\dot Q' + \f{3}{4} Q'' + 2Q = 0 ,\ee 
that is of course compatible with the original form (\ref{eqn:EOMQ}) once Eq.~(\ref{eqn:integrateout})
is backsubstituted.  The two methods are thus equivalent despite the fact that $Q$ is considered
an overdetermined variable in one and a propagating variable in the other.

Combining the EOM (\ref{ExampleE5}) with two consistency conditions $d\dot Q = \ddot Q dt + Q'' dr$ and $dQ'=\dot Q' dt + Q'' dr$, we obtain
\be  \label{EOMmat}
\begin{pmatrix}
1 & 2 & \f{3}{4} \\
dt & dr & 0 \\
0 & dt & dr
\end{pmatrix}
\begin{pmatrix}
\ddot Q \\ \dot Q' \\ Q''
\end{pmatrix}
=
\begin{pmatrix}
-2Q \\ d\dot Q \\ dQ'
\end{pmatrix} .
\ee
The $dt/dr$ values for which the matrix on the left hand side cannot be inverted  define the characteristic curves:
\be \f{dt}{dr} = 2,~ \f{2}{3} , \ee
which are consistent with those from the Kronecker form \eqref{SpecialKronecker}.

\subsubsection{Constrained higher order systems}

Our Kronecker analysis also assists in identifying cases where the EOMs appear to be of higher order
and require extra initial data or degrees of freedom but due to hidden constraints are really of a lower order \cite{Chen:2012au,Zumalacarregui:2013pma}.
For example, Ref.~\cite{Langlois:2015cwa} illustrate this phenomenon with the
coupled higher order DAE system for the fields $\{ \phi, q \}$
\begin{eqnarray}
a \ddddot \phi - k_0 \ddot \phi + b \dddot q  - c\ddot q - v \phi &=&0, \nonumber\\
k_1 \ddot q + b \dddot \phi + c \ddot \phi + w q &=& 0,
\label{eqn:higherorder}
\end{eqnarray}
where $\{ a, b, c, k_0, k_1, v, w \}$ are constants.
Since these equations are ODEs in time our ``matrix pencil" $\Amat +\lambda \Bmat=\Amat$.  Kronecker blocks in this case
can have no $\lambda$ and thus only
contain $\Lmat_0$, $\Lmat_0^P$ and $\Rmat_1(\infty)$.   Note that the Kronecker decomposition of $\Amat + \lambda \Cmat$ is
sometimes also used to decouple,  solve or study the stability  of fields in blocks in the DAE but we will not address that use here.

We perform our first order reduction
\begin{eqnarray} 
&& u_1 \equiv q , \nonumber\\
	{\rm EOM}_1&:& u_2 =  \dot u_1  (= \dot q) ,\nonumber\\
	{\rm EOM}_2&:& u_3 = \dot u_2   (= \ddot q) ,  \nonumber\\
&& u_4 \equiv \phi ,\nonumber\\
	{\rm EOM}_3&:&u_5 = \dot u_4 (= \dot \phi) ,  \nonumber\\
	{\rm EOM}_4&:&u_6 = \dot u_5 (=\ddot \phi) , \nonumber\\
	{\rm EOM}_5&:&u_7 = \dot u_6 (= \dddot \phi) ,
	\label{eqn:uidefs}
\end{eqnarray}
so that the original EOMs (\ref{eqn:higherorder}) are
\begin{eqnarray}
	{\rm EOM}_6&:& a \dot u_7 - k_0 u_6 + b \dot u_3 - c u_3 - v u_4 = 0, \nonumber\\
	{\rm EOM}_7&:&k_1 u_3 + b u_7 + c u_6 + w u_1 = 0 .
\end{eqnarray}
We therefore start with a $7 \times 7$ system with 5 defining equations and 2 original EOMs.

The original Kronecker structure is
\begin{eqnarray}
\{ \Lmat_0^P, 6 \times \Rmat_1(\infty) \}.
\end{eqnarray}
We can see immediately that the $\Lmat_0^P$ block is associated with EOM$_7$ 
which contains no derivatives.   
We use this to eliminate the highest order derivative
term $u_7$ assuming  $b\ne 0$
\begin{eqnarray}
u_7 &=& -\frac{k_1}{b} u_3 - \frac{c}{b}u_6 - \frac{w}{b} u_1,
\end{eqnarray}
which turns EOM$_6$ into 
\begin{eqnarray}
 (b^2 - a k_1) \mk{\dot u_3 -\frac{c}{b} u_3} &=&  \mk{b k_0 - \frac{a}{b} c^2 } u_6 + b v u_4
\nonumber\\
 && - \frac{a}{b} c w u_1 + a w u_2  ,
\end{eqnarray}
after using the definitional EOMs (\ref{eqn:uidefs}). 
For generic parameters, the resulting $6\times 6$ system is regular since this supplies the evolution equation for
$\dot u_3$.  As a whole, the evolution of
6 fields (3 phase space DOFs) are uniquely specified by initial values.

The special case is when $b^2 - a k_1=0$.  The system is singular since
there is no evolution equation for $u_3$.   This represents a column of zeros in the
$\Amat$ matrix or equivalently an $\Lmat_0$  underdetermined structure, but at the same
time we gain a constraint from EOM$_{6}$.  So the $6\times 6$ system is
\begin{eqnarray}
\{\Lmat_0, \Lmat_0^P, 5 \times \Rmat_1(\infty)  \}.
\end{eqnarray}
We can resolve the constraint by solving for the next highest derivative if $k_0 k_1-c^2\ne 0$
\begin{eqnarray}
 u_6 = \frac{ c w u_1 - b w u_2  -k_1 v u_4}{k_0 k_1-c^2},
 \label{eqn:u6}
\end{eqnarray}
which brings EOM$_4$ to
\begin{eqnarray}
	{\rm EOM}_4&:& \dot u_5 = \frac{ c w u_1 - b w u_2  -k_1 v u_4}{k_0 k_1-c^2}.
\end{eqnarray}
The constraint (\ref{eqn:u6}) also gives $\dot u_6$ which converts EOM$_5$ to another constraint
\begin{eqnarray}
u_3= k_1 \frac{c v u_4 + b v u_5 - k_0 w u_1}{k_0 k_1^2 - c^2 k_1 - b^2 w}.
\end{eqnarray}
Eliminating $u_3$ by assuming the denominator does not vanish brings EOM$_2$  to
\begin{eqnarray}
	{\rm EOM}_2&:& \dot u_2 = k_1 \frac{c v u_4 + b v u_5 - k_0 w u_1}{k_0 k_1^2 - c^2 k_1 - b^2 w} .
	\end{eqnarray}
The system is now a 
\begin{equation}
\{ 4 \times \Rmat_1(\infty) \} 
\end{equation} 
regular system of EOM$_{1-4}$ containing 4 hyperbolic blocks. All of the highest derivative field
have been eliminated by constraints hidden in the original EOMs.  We can of course also rewrite this as two coupled 
second order differential equations for $u_1$ and $u_4$, i.e.~the original $\phi$ and $q$.   The utility of this approach
for higher order systems is the algorithmic method of discovering constraints which in this case could have been done by
inspection.

\section{Decomposition techniques}
\label{sec:decomposition}

Here we review the decomposition techniques for regular or singular matrix pencils in \S\ref{sec:KroneckerAppendix} and for tensors on the 2-sphere in \S\ref{sec:harmonics}.  
In the main text and Appendix~\ref{sec:AlgorithmAppendix}, 
the former is used to block diagonalize and characterize the derivative structure of a set of partial differential equations of motion.  The latter is used to decouple the parity and angular
momentum modes of metric fluctuations.

\subsection{Kronecker decomposition of matrix pencil}
\label{sec:KroneckerAppendix}

Given two matrices $\Amat, \Bmat$ of the same dimensions, their linear combination $\Amat +
\lambda \Bmat$ is called a matrix pencil. 
If there exists a $\lambda$ for which $\Amat + \lambda \Bmat$ is invertible then it is called a regular pencil
and can be placed into a block diagonal Weierstrass form~\cite{Kagstrom:1983} of $r$ regular subblocks with invertible matrices $\Pmat$ and $\Qmat$
\ba
	{\Pmat(\Amat + \lambda \Bmat)\Qmat =}  \left\{ \Rmat_{\mu_1}, \ldots,
	\Rmat_{\mu_r} \right\} \nonumber.
\ea
As a shorthand convention, we denote a block diagonal concatenation of matrices $\diag\{\Amat_1,\ldots,\Amat_n\}$ as just the
list of its subblocks $\{ \Amat_1,\ldots, \Amat_n \}$.
Here each regular subblock $\Rmat_\mu$ is a $\mu \times \mu$ matrix pencil
defined by the generalized eigenvalue $\Omega$.  For finite $\Omega$
\be
	 \Rmat_\mu(\Omega) \equiv \lambda \Imat_\mu - \Jmat_\mu(\Omega),
\ee
where $\Imat_\mu$ is a $\mu \times \mu$ identity matrix and 
$\Jmat_\mu(\Omega)$ is a $\mu \times \mu$ lower Jordan block of the form
\be
\Jmat_\mu(\Omega) =\left(
\begin{array}{cccc}
\Omega &  & & \\
1 & \Omega &  & \\
& \ddots & \ddots & \\
& &  1 & \Omega
\end{array}
\right) ,
\label{eqn:Jordan}
\ee
whereas for the special case $\Omega=\infty$
\be
	 \Rmat_\mu(\infty) \equiv \lambda \Nmat_\mu - \Imat_\mu.
	 \label{eqn:Rnil}
\ee
Here $\Nmat_\mu=\Jmat_\mu(0)$ is a nilpotent lower Jordan matrix with $\Omega=0$.

If the matrix pencil is singular, it can still be cast into Kronecker form
\cite{VanDooren:1979aaa} with invertible matrices $\Pmat$ and $\Qmat$ 
\begin{eqnarray}
	\Pmat(\Amat + \lambda \Bmat)\Qmat &=& 
	\big\{\Lmat_{\mu_1},\ldots, \Lmat_{\mu_u},  \Lmat_{\mu_1}^P, \ldots, \Lmat_{\mu_o}^P, \nonumber\\
	&& \Rmat_{\mu_1}, \ldots, \Rmat_{\mu_r}\big\},
\end{eqnarray}
where $u$ is the number of ``underdetermined"   $\mu \times (\mu + 1)$  pencils of the form
\be
	\Lmat_\mu=\begin{pmatrix}
		\lambda & 1 & &\\
		& \ddots & \ddots &\\
		& & \lambda & 1
	\end{pmatrix} ,
\ee
and $o$ is the number of ``overdetermined" $(\mu + 1) \times \mu$  pencils of the pertransposed   $\Lmat_\mu$ form,
\be
	\Lmat_\mu^P =\begin{pmatrix}
		1 & &\\
		\lambda & \ddots &\\
		& \ddots & 1 \\
		& & \lambda
	\end{pmatrix} .
\ee
The degenerate  cases of $\Lmat_0$ and $\Lmat_0^P$ are formally $0 \times 1$ and $1 \times 0$ matrices
which stand for a  column or row of zeros  in the block diagonal form respectively.

Finally as a short hand convention, we denote  for example 
\be
\{ 4\times \Lmat_1 \} = \{ \Lmat_1,\Lmat_1,\Lmat_1,\Lmat_1 \},
\ee
if there are repeated identical block structures.

\subsection{Angular harmonics}
\label{sec:harmonics}

The normal modes or harmonic functions for tensorial fields on the 2-sphere are classified by their transformation properties
under a general rotation defined by Euler angles.  
Following Ref.~\cite{Goldberg} (see also \cite{Okamoto:2003zw}), we can decompose any trace free totally symmetric tensor 
of rank $s$
on the 2 sphere into its  spin $\pm s$ components ${}_{\pm s} f(\vh{n})$ as
\begin{equation}
T_{a_1\dots a_s}=({}_sf) \vp_{a_1}\dots\vp_{a_s} 
+({}_{-s}f )\vm_{a_1}\dots\vm_{a_s},
\label{Eqn:RankSBoth}
\end{equation}
where the covariant complex unit vectors on the sphere 
\begin{equation}
{ \vm}_a =\frac{1}{\sqrt{2}} \left(
\begin{array}{c}
1 \\
i\sin\theta\\
\end{array} 
\right),
\quad
{\vp}_a =\frac{1}{\sqrt{2}} \left(
\begin{array}{c}
1 \\
-i\sin\theta\\
\end{array} 
\right),
\end{equation}
obey the conjugate orthonormality property
\begin{eqnarray}
\vm_a  \vm^a  = \vp_a \vp^a = 0 ,\qquad  \vm_a  \vp^a  = 1.
\label{Eqn:Orthonormality}
\end{eqnarray}
Angular indices are raised and lowered by the metric $\smetric_{ab}$ on the 2-sphere
and the antisymmetric Levi-Civita tensor $\epsilon_{ab}$  converts the real and imaginary parts 
\begin{eqnarray}
\epsilon_{a}^{\hphantom{a}b} \vm_b &=&  i \vm_a, \nonumber\\
\epsilon_{a}^{\hphantom{a}b} \vp_b &=& - i \vp_a.
\end{eqnarray}
Explicitly,
\begin{equation}
\smetric_{ab} = \left(
\begin{array}{cc}
1 & 0\\
0 & \sin^2 \theta\\
\end{array} 
\right) ,\quad
\epsilon_{ab} = \left(
\begin{array}{cc}
0 & \sin\theta \\
-\sin\theta & 0\\
\end{array} 
\right) .
\end{equation}
Note that $\vm_a \vp_b + \vp_a \vm_b = \smetric_{ab}$.
In this Appendix, we employ $\vh{n}$ to denote the radial unit vector specified by
the angular coordinates $\{ \theta,\phi \}$ and integrals over 
$d\vh{n}$ as integrals over angles on the 2-sphere.  A right handed rotation of the coordinate axis 
around $\vh{n}$ by $\psi$ changes the spin functions by a phase $e^{-i s\psi}$.
These definitions apply to $s=0$ scalar functions as well but note that
Eq.~(\ref{Eqn:RankSBoth}) implies the convention $T = {}_0 f + {}_{-0} f = 2 {}_0 f$.  In this case the complete
set of modes for $T$ are the spherical harmonics $Y_{\ell m}$.

The spin-$s$ functions can likewise be decomposed into multipole moments based on their
transformation properties under the remaining Euler angles, i.e.~a rotation of the pole of the spherical coordinates.
The normal modes are generalizations of spherical harmonics called spin spherical harmonics 
\cite{Goldberg} 
that obey 
the orthonormality  property
\begin{equation}
\int d\vh{n} ( {}_s Y_{\ell' m'}^* )( {}_s Y_{\ell m} )= \delta_{\ell \ell'}\delta_{m m'},
\end{equation}
the conjugation property
\begin{equation}
{}_s Y_{\ell m}^* = (-1)^{m+s} {}_{-s}Y_{\ell (-m)},
\end{equation}
and the parity property
\begin{equation}
{}_s Y_{\ell m}(\vh{n})  =  (-1)^{\ell} {}_{-s}Y_{\ell m}(-\vh{n})
\end{equation}
where $\ell \ge s$ and $-\ell \le m \le \ell$.
Rotation of the coordinate origin mixes the $m$ moments of a given angular momentum $\ell$.

Thus the $s\ge 1$ tensor  eigenstates of a given angular momentum $(\ell,m)$ with even parity $X=E$ and odd parity $X=B$
are  given by 
\begin{equation}
	Y^X_{\ell m,a_1 \ldots a_s} ={}_s f^X_{\ell m} \vp_{a_1}\ldots \vp_{a_s}  + {}_{-s}  f^X_{\ell m} \vm_{a_1} \ldots \vm_{a_s},
\end{equation}
where the spin functions are
\begin{eqnarray}
{}_s f^E_{\ell m} &=& -i( {}_s f^B_{\ell m}) =  \frac{{}_{s}Y_{\ell m}}{\sqrt{2} } (-1)^s, \nonumber\\
{}_{-s} f^E_{\ell m}&=& i( {}_{-s} f^B_{\ell m}) = \frac{{}_{-s}Y_{\ell m}}{\sqrt{2} }.
\end{eqnarray}
By virtue of the analogous spin relations above,
the tensors satisfy the orthonormality relation
\begin{eqnarray}
\int d\vh{n} Y^{X*}_{\ell m,a_1 \ldots a_s} Y^{X',a_1 \ldots a_s}_{\ell' m'} = \delta_{\ell\ell'}\delta_{m m'}\delta_{X X'}
\label{eqn:orthonormality}
\end{eqnarray}
and the conjugation relation
\begin{eqnarray}
Y^{X*}_{\ell m,a_1 \ldots a_s} = (-1)^m Y^{X}_{\ell (-m),a_1 \ldots a_s} ,
\end{eqnarray}
where $X \in E,B$.

Covariant differentiation on these tensors raise and lower the spin weights according to the
ladder operators $\edth,\baredth$ \cite{Goldberg} 
\begin{eqnarray}
\nabla_b T_{a_1\ldots a_s} &=& -\vp_{a_1}\cdots\vp_{a_s} 
\frac{\vp_b  \edth 
+\vm_b \baredth }{\sqrt{2}}
{}_sf \nonumber\\
&&-  \vm_{a_1}\cdots\vm_{a_s}
\frac{\vp_b  \edth 
+\vm_b \baredth }{\sqrt{2}}{}_{-s}f .
\label{Eqn:CovDerivFinal}
\end{eqnarray}
 In particular, their action on the spin harmonics gives
 \begin{eqnarray}
\edth {}_s Y_{\ell m} &=& \sqrt{(\ell-s)(\ell + s +1)} \, {}_{s+1} Y_{\ell m} ,\nonumber\\
\baredth  {}_s Y_{\ell m} &=&- \sqrt{(\ell+s)(\ell - s +1)}\,  {}_{s-1} Y_{\ell m}.
\end{eqnarray}

To make a connection with the RWZ literature, we can use Eq.~(\ref{Eqn:CovDerivFinal}) 
to  relate the covariant derivative of the scalar harmonics to the vector harmonics
\ba
	Y^E_{\ell m,a} = \frac{ \nabla_a Y_{\ell m} }{\sqrt{\ell(\ell+1)}}, \quad 
	Y^B_{\ell m,a} = \frac{\epsilon_{ba} \nabla^b Y_{\ell m}}{\sqrt{\ell(\ell+1)}}  , 
\ea
and likewise the second derivative to the rank-2 tensor harmonics \cite{Kamionkowski:1996ks}
\ba
	Y^E_{\ell m,ab} &=& \sqrt{2\frac{(\ell-2)!}{(\ell+2)!}}
	 \(\nabla_a \nabla_b -
	\frac{1}{2}\smetric_{ab} \nabla_c \nabla^c\)Y_{\ell m}, \nonumber\\
	Y^B_{\ell m,ab} &=& \sqrt{\frac{1}{2}\frac{(\ell-2)!}{(\ell+2)!}}
	\(\epsilon^c_{\phantom{c}b}\nabla_a \nabla_c +
	\epsilon^c_{\phantom{c}a}\nabla_b \nabla_c \)Y_{\ell m}\nonumber .
\ea

Eq.~(\ref{Eqn:CovDerivFinal})  also  provides identities for integrals of scalar contractions of 
covariant derivatives of tensors over the 2-sphere
\begingroup
\allowdisplaybreaks[1]
\begin{eqnarray}
&& \int d\vh{n}(\nabla_b Y^{X*}_{\ell m,a_1 \ldots a_s})( \nabla^b Y^{X',a_1 \ldots a_s}_{\ell' m'} )
\nonumber\\*
&&\qquad = \delta_{\ell\ell'}\delta_{m m'}\delta_{X X'}[\ell(\ell+1)-s^2] ,\nonumber\\
&& \int d\vh{n}(\nabla_{a_s}  Y^{X*}_{\ell m,a_1 \ldots a_{s-1}}) Y^{X',a_1 \ldots a_s}_{\ell' m'} 
\nonumber\\*
&&\qquad = \delta_{\ell\ell'}\delta_{m m'} \delta_{X X'}  \sqrt{\frac{(\ell-s+1)(\ell+s)}{2}},
\nonumber\\
&& \int d\vh{n}(\nabla_{b}  Y^{X*}_{\ell m,a_1 \ldots a_{s}})( \nabla^{a_s} Y^{X',a_1 \ldots a_{s-1}b}_{\ell' m'} )
\nonumber\\*
&&\qquad = \delta_{\ell\ell'}\delta_{m m'}\delta_{X X'}\frac{(\ell-s)(\ell+s+1) }{2},
\nonumber\\
&& \int d\vh{n}(\nabla_bY^{X*}_{\ell m,a_1 \ldots a_{s}})(  \nabla_c  Y^{X',a_1 \ldots a_s b c}_{\ell' m'} )
\nonumber\\*
&&\qquad = -\delta_{\ell\ell'}\delta_{m m'}\delta_{X X'}\sqrt{\frac{(\ell -s)(\ell+s+1)}{2}}
\nonumber\\*
&&\qquad \quad \times \sqrt{\frac{(\ell+s+2)(\ell-s-1)}{2}} ,
\label{eqn:angularidentities}
\end{eqnarray}
\endgroup
which are used in the main text to determine how the various spin components are coupled through the equations
of motion.

\vfill

\begin{widetext}
\section{Alternative odd analysis}
\label{ssec:oddal}

In this appendix we highlight the difference between the odd mode analysis in \S\ref{ssec:oddl2}
and an alternative analysis employing a technique of auxiliary fields, which is commonly
used in the literature~\cite{DeFelice:2011ka, Motohashi:2011pw, Motohashi:2011ds,
Kobayashi:2012kh, Ogawa:2015pea}.  
See also Appendix~\ref{sec:proderi} for a simpler example of the general technique.

The general $B$ mode Lagrangian \eqref{OddLagr} is given explicitly by 
\ba
	 \mathcal{L}_{B} &=& D_1 h_0^2 + D_2 h_1^2 + D_3 h_2^2 + D_4 h_0 h_1
	+ D_5 h_0 h_2
 + D_6 h_1 h_2+ D_7 (\dot h_1 - h_0')^2 + D_8 \dot h_2^2 + D_9
	h_2'^2\nonumber\\
	&&
 + D_{10} h_1 h_0' + D_{11} h_1 h_2' + D_{12} h_0 \dot h_1 +
	D_{13} h_0 \dot h_2 ,
		\label{OddLagrRepeat}
\ea
where the $D_i$ coefficients in terms of the background metric and \stucky fields are
\ba
D_1 &=&\frac{a^2 \left[ 3 r^2 \dot a^2-2 r b'
b+ \Lnorm  r^2 b^2  \OurG_{22}+ \ell
(\ell+1)b^2\ \right]
+2 r a' a b \left(b-rb'\right)
+ r^2 a'^2 b^2-\Leff  r^2 a^4
b^2}{4 r^2 a^3 b^3}
, \nonumber\\
D_2 &=&\frac{- a^2 \left[ r^2 \dot a^2-2 r b'
b+(\ell-1)(\ell+2) b^2\right]+2 r a' a b \left(r
b'+b\right)+ r^2 a'^2 b^2+ r^2 a^4 \left(\Leff b^2+ \Lnorm  \OurG_{11}\right)}{4 r^2
a^5 b}
, \nonumber\\
D_3 &=&\frac{-a^2 \left[4 r^2 \dot a^2+b^2
\left(\Lnorm  r^2  \OurG_{22}-2\right)\right]+4 r^2
a'^2 b^2+8 r a' a b^2+ r^2 a^4 \left(2 \Leff  b^2+  \Lnorm  \OurG_{11}\right)}{16 r^4
a^5 b}
,
\nonumber\\
D_4 &=&-\frac{ 4 \dot a (r a'+a)
+\Lnorm  r  a^2 \OurG_{12}}{2 r
a^3 b}
, \quad
D_5 =-\frac{\sqrt{(\ell-1)(\ell+2)} \dot a}{2 r^2 a^2 b}
, \quad 
D_6 =\frac{\sqrt{(\ell-1)(\ell+2)} \left(r a'+a\right)b}{2
r^3 a^4}
, \nonumber\\
D_7 &=&\frac{1}{4 a b}
, \quad
D_8 =\frac{1}{16 r^2 a b}
, \quad
D_9 =-\frac{ b}{16 r^2 a^3}
, \quad
D_{10} =\frac{\dot a}{a^2 b}
, \quad
D_{11} =-\frac{\sqrt{(\ell-1)(\ell+2)} b}{4 r^2 a^3}
, \quad
D_{12} =\frac{ \left(r a'+a\right)}{r a^2 b}
, \nonumber\\
D_{13} &=&\frac{\sqrt{(\ell-1)(\ell+2)}}{4 r^2 a b}
.
\label{eqn:Ds}
\ea
Parts proportional to $\Lnorm$ are contributions from the dRGT potential term in the
quadratic Lagrangian \eqref{SimplifiedLagr}. Naturally, this potential term affects only the coefficients of nonderivative
terms $D_1, D_2, D_3$, and $D_4$. The remaining parts of the coefficients $D_i$ are
inherited from the Einstein-Hilbert Lagrangian and the effective cosmological constant 
$\Lambda$.

While we cannot integrate out either $h_0$ or $h_1$
from the Lagrangian \eqref{OddLagr} directly, the peculiar structure $(h_0' - \dot
h_1)^2$ enables us to introduce an auxiliary field and obtain a dynamically
equivalent unconstrained
Lagrangian with only two dynamical field variables.  
The first step is to complete the square of derivative terms of $h_0$ and $h_1$ in Eq.~\eqref{OddLagr} as
\be
	\mathcal{L}_{B} = D_7 \( \dot h_1 - h_0' + \frac{D_{12}}{2 D_7}h_0 - \frac{D_{10}}{2
	D_7} h_1\)^2+\cdots .
\label{csqLagr}
\ee
Next introduce an auxiliary field $q$ defined to be
\be
	\mathcal{L}_{B} = -\f{\qa^2}{D_7} + 2 \qa \(\dot h_1 - h_0' + \frac{D_{12}}{2 D_7}h_0 - \frac{D_{10}}{2 D_7} h_1\) +\cdots.
\label{QLagr}
\ee
Clearly, the equation of motion for $q$ is given by
\be
\label{qRelation}
	\qa = D_7 \mk{ \dot h_1 - h_0' + \frac{D_{12}}{2 D_7}h_0 - \frac{D_{10}}{2 D_7} h_1 } ,
\ee
and we recover Eq.~\eqref{csqLagr} by plugging Eq.~\eqref{qRelation} back into Eq.~\eqref{QLagr}.  
In this way, the problematic $(h_1' - \dot h_0)^2$ term in Eq.~\eqref{OddLagr} can be effectively hidden inside the
terms of Eq.~\eqref{QLagr}, while the remaining derivatives on $h_0, h_1$ can be moved
onto $\qa$ through integration by parts. After all the algebra, we get a
Lagrangian without derivatives on $h_0, h_1$
\ba
\label{AfterHayatoTrick}
	\mathcal{L}_{B} &=& \hat D_1 h_0^2 + \hat D_2 h_1^2 + D_3 h_2^2 +
	\hat D_4 h_0 h_1 + D_5 h_0 h_2
  + D_6 h_1 h_2+ D_8 \dot h_2^2 + D_9 h_2'^2 + D_{11} h_1 h_2'  + D_{13} h_0 \dot h_2 \label{Loddq}
	\\
	&&~ - \f{1}{D_7} \qa^2 + \f{D_{12}}{D_7} \qa h_0 - \f{D_{10}}{D_7} \qa h_1 -  2\dot \qa h_1 + 2 \qa' 
	h_0 , \nonumber
\ea
where the equal sign should be interpreted as the same up to boundary terms from integrating the respective Lagrangians by parts.
Due to the integrations by parts
\ba
	\hat D_1 &=& D_1 - \frac{D_{12}^2}{4D_7} - \frac{D_{12}'}{2} 
	=  \f{(\ell-1)(\ell+2) + \Lnorm  r^2 \bar\chi_{22} }{4 r^2 a b}, \nonumber\\
	\hat D_2 &=& D_2 - \frac{D_{10}^2}{4D_7} - \frac{\dot D_{10}}{2} 
	=  \f{-(\ell-1)(\ell+2)b^2 +  \Lnorm  r^2  a^2 \bar\chi_{11} }{4 r^2 a^3 b} ,\nonumber\\
\nonumber \\
	\hat D_4 &=& D_4 + \frac{D_{12}D_{10}}{2D_7} = - \f{ \Lnorm  \bar\chi_{12} }{2 ab} .
		\label{hatDs}
\ea
Notice that the case of $\ell=1$ is special;
for this reason, we consider $\ell =1$ and $\ell \ge 2$ separately.

\subsection{Odd $\ell \ge 2$ EOMs}

For modes with $\ell \ge 2$, 
we can integrate $h_0, h_1$ out by using their equations of motion. 
The end result reads
\ba
	\label{HayatosSolution}
	h_0 &=& -\frac{\hat D_4 [-2 \dot \qa + D_6 h_2 + D_{11} h_2' - \f{D_{10}}{D_7}\qa] -
	2 \hat D_2 \left[2 \qa' + D_{13} \dot h_2 + D_5 h_2 + \f{D_{12}}{D_7}\qa \right]}{\hat D_4^2 - 4 \hat
	D_1 \hat D_2} ,
	\nonumber\\
	h_1 &=& - \frac{2 \hat D_1 [ 2 \dot \qa - D_6 h_2 - D_{11} h_2' + \f{D_{10}}{D_7}\qa]
	+ \hat D_4 \left[2 \qa' + D_{13}\dot h_2 +D_5 h_2 + \f{D_{12}}{D_7}\qa \right]}{\hat D_4^2 - 4 \hat
	D_1 \hat D_2} .
\ea
The coefficient in the denominator is
\ba 
	{\hat D_4^2 - 4 \hat D_1 \hat D_2 = \frac{(\ell-1)(\ell+2)}{4 r^4  a^4}}
	 \left[
	(\ell-1)(\ell+2) + \Lnorm r^2  a^2 \Tr \OurG\right] , \label{h0h1det}
\ea
and is typically nonzero for $\ell \geq 2$.
However, there are positions in spacetime
where \eqref{h0h1det} vanishes and we cannot solve for $h_0, h_1$ through 
Eq.~\eqref{HayatosSolution}. 
Because $\Tr \OurG$ is an invariant quantity, we cannot avoid this problem by
going into another slicing of the background spacetime as we did in our main analysis.
Outside of these problematic points, by plugging solutions \eqref{HayatosSolution} into
the Lagrangian \eqref{AfterHayatoTrick} it is possible to obtain an unconstrained
Lagrangian with only two degrees of freedom $\qa, h_2$ and with no more than second
derivatives.  This Lagrangian then leads to two second order equations of motion for
$\qa,h_2$. Characteristic curves for these EOMs can be then found in a standard way
\cite{Hoffman:2001aaa} by focusing only on the second derivative terms in the two equations of
motion and determining where the EOMs fail to determine their values given the lower derivatives.
Similarly to \eqref{EOMmat}, requiring four consistency conditions such as
\be
	d(\qa') = \qa'' dr + \dot \qa' dt,  
\ee
where the left hand side is assumed to be continuous, leads to a linear system of six equations
for the six unknown second derivatives $\qa'', \dot \qa', \ddot \qa, h_2'', \dot h_2', \ddot
h_2$. For general values of the infinitesimal displacement vectors $dt, dr$ this system
has a unique solution. For special ratios $dt/dr$ which correspond to the characteristic
curves the system allows for multiple solutions and the highest derivatives are not
uniquely defined. Characteristic curves obtained this way agree with the curves obtained
by our main analysis.
Note that this alternate procedure does not on its own distinguish between two $\Rmat_1$ subsystems
which share characteristics and the $\Rmat_2$ system identified in our main analysis.  Likewise
since the discontinuity identified here is only in the highest derivatives, the analysis 
does not address the chained derivatives in the $\Rmat_2$ block that
link characteristics.

\subsection{Odd $\ell=1$ EOMs}
\label{sec:oddl1app}

For $\ell=1$, there is no spin-2 $h_2$ mode 
so that Eq.~\eqref{Loddq} becomes
\begin{align} 
\label{Loddl1q}
\mathcal L_{B,\ell=1} =& \hat D_1 h_0^2 + \hat D_2 h_1^2 + \hat D_4 h_0 h_1 - \f{\qa^2}{D_7} + \f{D_{12}}{D_7} \qa h_0
- \f{D_{10}}{D_7} \qa h_1 - 2 \dot \qa h_1 + 2 \qa' h_0.  
\end{align}
Note that the coefficients $\hat D_1,\hat D_2,\hat D_4$
in Eq.~\eqref{hatDs} are proportional to $m^2$ for $\ell=1$, which makes the following analysis different from general relativity.
Variation of Eq.~\eqref{Loddl1q} with respect to $h_0, h_1, \qa$ yields 
\begin{align}
2 \hat D_1 h_0 + \hat D_4 h_1 + \f{D_{12}}{D_7} \qa + 2\qa' &=0, \label{ell1eoms1} \\
\hat D_4 h_0 + 2 \hat D_2 h_1 - \f{D_{10}}{D_7} \qa - 2\dot \qa &=0, \label{ell1eoms2}\\
- 2 \qa + D_{12} h_0 - D_{10} h_1 + 2 D_7 \dot h_1 - 2 D_7 h_0' &=0. \label{ell1eoms3}
\end{align}
Since $\hat D_4^2-4\hat D_1\hat D_2=0$ for $\ell=1$ from Eq.~(\ref{h0h1det}), we cannot solve Eqs.~\eqref{ell1eoms1} and \eqref{ell1eoms2} for $h_0$
and $h_1$.  Instead 
we first solve \eqref{ell1eoms1} for $h_0$:
\be h_0 = -\f{1}{2\hat D_1} \mk{ \hat D_4 h_1 + \f{D_{12}}{D_7}\qa + 2\qa' } . \label{eqh0}  \ee
Plugging Eq.~\eqref{eqh0} into Eq.~\eqref{ell1eoms2}, we obtain an autonomous equation 
\be \dot \qa + \f{\hat D_4}{2 \hat D_1} \qa' + \f{1}{2 D_7} \mk{ D_{10} + \f{\hat D_4 D_{12}}{2\hat D_1} } \qa = 0 .  \label{eqq} \ee
Here, by the virtue of $\hat D_4^2-4\hat D_1\hat D_2=0$, the $h_1$ term drops out.
Finally, from Eq.~\eqref{ell1eoms3} we obtain
\begin{align} 
&\dot h_1 + \f{\hat D_4}{2\hat D_1} h_1' + \kk{ \mk{\f{\hat D_4}{2\hat D_1}}' - \f{1}{2D_7} \mk{D_{10} + \f{\hat D_4 D_{12}}{2\hat D_1} }  } h_1 
= - \mk{\f{ \qa'}{\hat D_1}}' +  \kk{ \f{1}{D_7} + \f{D_{12}^2}{4D_1D_7^2} - \mk{\f{D_{12}}{2D_1D_7}}' } \qa. \label{eqh1}  
\end{align}
Note that the source term in the right-hand side is written in terms of $\qa$.  Therefore,
given background evolution, we can first solve Eq.~\eqref{eqq} for $\qa(t,r)$, plug it into
Eq.~\eqref{eqh1} to solve for $h_1(t,r)$, and then Eq.~\eqref{eqh0} gives $h_0(t,r)$.     As we
have two first-order differential equations, we require two initial conditions to solve
the system. It is straightforward from Eq.~\eqref{eqq} and \eqref{eqh1} to check that 
the characteristic curves corresponding to
these two equations are the same as those uncovered in our main analysis.

A structurally similar set of equations was uncovered for the two $\ell=0$
modes in isotropic gauge~\cite{Wyman:2012iw, Motloch:2015gta}. There one of the isotropic modes $\delta
\Gamma$ formed an autonomous equation; this mode then sourced the second
isotropic mode $\delta f$. Unlike here, both these equations were manifestly first order,
with $\delta \Gamma$ sourcing $\delta f$ through a term without any derivatives. In the
present analysis we see $h_1$ sourced by up to second derivatives of $\qa$. Because of
this derivative sourcing, this system is an $\Rmat_2$ parabolic block whereas $\ell = 0$
is a $2\times \Rmat_1$ pair of hyperbolic blocks.

\subsection{Odd $\ell=1$ Hamiltonian analysis}
\label{ssec:oddl1Hamiltonian}

The $\ell=1$ odd Lagrangian is simple enough to also perform the Hamiltonian analysis.  
From \eqref{Loddl1q}, the canonical momenta for $\qa$, $h_0$, $h_1$ are given by
\be p_q = -2h_1, \quad p_0 = 0, \quad p_1 = 0, \ee
and yield three primary constraints:
\be \phi_q = p_q + 2h_1, \quad \phi_0 = p_0, \quad \phi_1= p_1. \ee
The only nonvanishing Poisson bracket between them is 
\be \{ \phi_q,\phi_1 \} =2. \ee
The total Hamiltonian density is given by
\begin{align} 
{\cal H}_T =& \dot \qa p_q + \dot h_0 p_0 + \dot h_1 p_1 - {\cal L}_{B,\ell=1} + \mu_q \phi_q + \mu_0 \phi_0 + \mu_1 \phi_1, \notag\\ 
=& - \hat D_1 h_0^2 - \hat D_2 h_1^2 - \hat D_4 h_0 h_1 +\f{1}{D_7} \qa^2 - \f{D_{12}}{D_7} \qa h_0
+ \f{D_{10}}{D_7} \qa h_1 - 2\qa' h_0 + \mu_q \phi_q + \mu_0 \phi_0 + \mu_1 \phi_1 ,
\end{align}
where $\mu_q$, $\mu_0$, $\mu_1$ are Lagrange multipliers.
The consistency conditions are then given by
\begin{align}
&0\approx \dot \phi_q = \{ \phi_q,{\cal H}_T \} = -\f{2}{D_7}\qa + \f{D_{12}}{D_7}h_0 - \f{D_{10}}{D_7}h_1 + 2\mu_1,\notag\\
&0\approx \dot \phi_0 = \{ \phi_0,{\cal H}_T \} = 2\hat D_1h_0 + \hat D_4h_1 + \f{D_{12}}{D_7}\qa + 2\qa', \notag\\
&0\approx \dot \phi_1 = \{ \phi_1,{\cal H}_T \} = 2\hat D_2h_1 + \hat D_4h_0 - \f{D_{10}}{D_7}\qa - 2\mu_q.
\end{align}
From $\dot \phi_q\approx 0$ and $\dot \phi_1\approx 0$ we can solve for $\mu_1$ and $\mu_q$, 
while from $\dot \phi_0\approx 0$ we obtain a secondary constraint
\be \phi_2 = 2\hat D_1h_0 + \hat D_4h_1 + \f{D_{12}}{D_7}\qa + 2\qa'. \ee
Poisson brackets of $\phi_2$ with the remaining constraints are
\begin{align}
&\{ \phi_2, \phi_q \} = \f{D_{12}}{D_7} + 2\{ \qa',p_q \}, \notag\\
&\{ \phi_2, \phi_0 \} = 2\hat D_1, \notag\\
&\{ \phi_2, \phi_1 \} = \hat D_4.
\end{align}
Therefore, the consistency condition of $\phi_2$ gives a relation between $\mu_q$, $\mu_0$, $\mu_1$
and does not generate a further constraint.  
The determinant of the Poisson brackets between constraints is given by 
\be \det \{\phi_i ,\phi_j\} = 16\hat D_1^2. \ee
So long as $\hat D_1 = \Lnorm  \OurG_{22}/(4 a b) \not = 0$, all four constraints are second class.
Therefore, the number of initial conditions we need is $3\times 2-4=2$, which is consistent with two
first-order EOMs for $\qa$ and $h_1$ obtained above.

Finally by using the constraints $\phi_i=0$, we can express
\be
	h_1 = -\frac{p_q}{2}, \quad	   h_0 = \frac{\hat D_4 p_q/2 - D_{12}q/D_7 -2 q'}{2\hat D_1} ,
\ee
and rewrite the Hamiltonian on the constrained surface in terms of $q, p_q$
\be
	 {\cal H}_T = - \f{(2\hat D_1 D_{10} + \hat D_4 D_{12})q + 2\hat D_4 D_7 q'}{4\hat D_1D_7} p_q
	+ \f{D_{12}^2 + 4\hat D_1 D_7}{4\hat D_1 D_7^2} q^2
	+ \f{D_{12}}{\hat D_1 D_7} q q' + \f{1}{\hat D_1} q'^2 .
\ee
The quadratic term
$p_q^2$ vanishes because of $\hat D_4^2 - 4 \hat D_1 \hat D_2 = 0$.  
This Hamiltonian is linear in $p_q$ and thus unbounded from below.

\end{widetext}
\bibliography{reggew}

\end{document}